\def\singlecol{0}

\if\singlecol0
\documentclass[journal]{IEEEtran}
\else
\documentclass[12pt,draftclsnofoot,onecolumn]{IEEEtran}
\fi

\def\editmode{0}

\def\bibfilenames{WISENET}

\if\editmode1  
\usepackage[backend=bibtex,style=alphabetic,sorting=debug]{biblatex}
\DeclareFieldFormat{labelalpha}{\thefield{entrykey}}
\DeclareFieldFormat{extraalpha}{}
\bibliography{\bibfilenames}
\newcommand{\cmt}[1]{\noindent\textcolor{lightgreen}{\underline{[#1]}}} 
\newcommand{\hc}[1]{\textcolor{blue}{#1}} 
\newenvironment{myitemize}{\begin{itemize}}{\end{itemize}}
\newcommand{\myitem}{\item}

\else
\usepackage{cite}
\usepackage[skip=0pt]{caption}
\usepackage{flushend}
\newcommand{\cmt}[1]{} 
\newcommand{\hc}[1]{\textcolor{black}{#1}} 
\newenvironment{myitemize}{}{}
\newcommand{\myitem}{}

\fi 

\usepackage{fixltx2e}

\usepackage{graphicx}

\usepackage{subcaption}              

\usepackage[utf8]{inputenc}

\usepackage{amsfonts}
\usepackage{amsmath}
\usepackage{mathtools}

\usepackage{amssymb}

\usepackage{bm}

\usepackage{color,verbatim}
\usepackage{multirow}
\usepackage{accents}

\usepackage{theoremref}

\usepackage{url}

\newcounter{rulecounter}
\newcommand{\resetrule}{ \setcounter{rulecounter}{0}}
\resetrule

\newsavebox{\selvestebox}
\newenvironment{colbox}[1]
  {\newcommand\colboxcolor{#1}%
   \begin{lrbox}{\selvestebox}%
   \begin{minipage}{\dimexpr\columnwidth-2\fboxsep\relax}}
  {\end{minipage}\end{lrbox}%
   \begin{center}
   \colorbox{\colboxcolor}{\usebox{\selvestebox}}
   \end{center}}

\definecolor{orange}{rgb}{1,0.8,0}
\definecolor{gray}{rgb}{.9,0.9,0.9}
\definecolor{darkgray}{rgb}{.3,0.3,0.3}
\definecolor{darkblue}{rgb}{.1,0.0,0.3}
\definecolor{lightblue}{rgb}{0.7,0.7,1}
\definecolor{lightred}{rgb}{1,0.7,.7}
\definecolor{purple}{RGB}{204,153,255}
\definecolor{lightgray}{rgb}{.95,0.95,0.95}
\definecolor{lightgreen}{rgb}{0.3,0.5,0.3}
\definecolor{darkgreen}{rgb}{0.05,0.3,0.05}

\newcommand{\ra}{$\rightarrow$~}

\newcommand{\brackets}[1]{\left\{#1\right\}}

\newcommand{\tbm}[1]{{\tilde{\bm #1}}}

\newcommand{\hbm}[1]{{\hat{\bm #1}}}

\newcommand{\rfield}{\mathbb{R}}

\newcommand{\transpose}{^T}
 \newcommand{\define}{\triangleq}

\newcommand{\expected}[1]{\mathop{\textrm{E}}\brackets{#1} }

\newcommand{\minimize}{\mathop{\text{minimize}}}

\newtheorem{myproposition}{Proposition}
\newtheorem{myremark}{Remark}
\newtheorem{myproblemstatement}{Problem Statement}
\newtheorem{mylemma}{Lemma}
\newtheorem{mytheorem}{Theorem}
\newtheorem{mydefinition}{Definition}
\newtheorem{mycorollary}{Corollary}

\newcommand{\firstrev}[1]{\textcolor{black}{#1}}
\newcommand{\secondrev}[1]{\textcolor{black}{#1}}
\newcommand{\thirdrev}[1]{\textcolor{black}{#1}}
\newcommand{\fourthrev}[1]{\textcolor{black}{#1}}

\begin{document}
\renewcommand{\define}{{:=}}
\renewcommand{\transpose}{^\top} 

\newcommand{\dataset}{{\hc{\mathcal{D}}}}

\newcommand{\sourcenum}{{\hc{ S}}}
\newcommand{\sourceind}{{\hc{ s}}}
\newcommand{\basenum}{{\hc{ B}}}
\newcommand{\baseind}{{\hc{ b}}}
\newcommand{\bases}{{\hc{\beta}}}
\newcommand{\truecoeffs}{{\hc{\pi}}}
\newcommand{\estimatecoeffs}{{\hc{\hat \pi}}}
\newcommand{\truecoeffsmat}{{\hc{\bm \Pi}}}
\newcommand{\estimatecoeffsmat}{{\hc{\hbm \Pi}}}
\newcommand{\transmitpsd}{{\hc{\Upsilon}}}
\newcommand{\measpsd}{{\hc{\tilde \truepsd}}}
\newcommand{\estimatepsd}{{\hc{\hat \truepsd}}}
\newcommand{\estimatepsdmat}{{\hc{\hbm \truepsd}}}

\newcommand{\measpsdmat}{{\tbm \truepsd} }
\newcommand{\augmeaspsdmat}{{\hc{\textcolor{black}{\breve{\bm \Psi}}}}}
\newcommand{\gridpoint}{{\hc{\bm \xi}}}
\newcommand{\nnfun}{\hc{p_{\bm w}}}
\newcommand{\nnbemfun}{\hc{\bar p_{\bm w}}}

\newcommand{\truepsd}{\hc{\Psi}}
\newcommand{\mask}{\hc{M}}
\newcommand{\metamap}{\hc{\mask'}}
\newcommand{\building}{\hc{\mathcal{B}}}
\newcommand{\samplingfactor}{{\hc{ \gamma}}}

\newcommand{\pars}{\hc{\bm w}}
\newcommand{\encfun}{\hc{\epsilon}_{\pars}}
\newcommand{\decfun}{\hc{\delta}_{\pars}}
\newcommand{\code}{\hc{\bm \lambda}}
\newcommand{\entrycode}{\hc{\lambda}}
\newcommand{\latentnum}{\hc{N_\lambda}}
\newcommand{\latentvarmat}{\hc{ \bm \Lambda}}
\newcommand{\latentcovmat}{\hc{ \bm C_{\lambda}}}

\newcommand{\auxinmat}{{\hc{\bm\Phi}}}
\newcommand{\auxlayerin}{\auxinmat^{(I)}}
\newcommand{\auxlayerout}{\auxinmat^{(O)}}
\newcommand{\inchind}{{\hc{c_\text{in}}}}
\newcommand{\inchnum}{{\hc{C_\text{in}}}}
\newcommand{\outchind}{{\hc{c_\text{out}}}}
\newcommand{\outchnum}{{\hc{C_\text{out}}}}

\newcommand{\layerfun}{\hc{p}}
\newcommand{\layerind}{{\hc{l}}}
\newcommand{\layernum}{\hc{L}}
\newcommand{\layerparnot}[2]{^{\hc{(}#1\hc{)}}_{#2}}
\newcommand{\psdvarmat}{\hc{\bm \chi}}

\title{Deep Completion Autoencoders for \\
Radio Map Estimation}
\author{Yves Teganya$^{1}$ and Daniel Romero$^2$\\
 $^1$ Ericsson Research, 1640 40 Kista, Sweden.
  \\
  $^2$ Dept. of Information and Communication Technology, University
  of Agder, 4879 Grimstad, Norway.
  \thanks{
Emails: daniel.romero@uia.no and yves.teganya@ericsson.com. 
This work was supported by the Research Council of Norway through the FRIPRO TOPP-FORSK Grant 250910/F20 and the IKTPLUSS Grant \fourthrev{311994.} 
Parts of this work were be presented at the IEEE International
 Conference on Communications 2020~\cite{teganya2020autoencoders}. This
 work took place while the first author was with the University of
 Agder, Norway.
 }
}



\maketitle
\begin{abstract}
Radio  maps provide metrics such as power spectral density for every
location in a geographic area and find numerous applications such as
UAV communications, interference control, spectrum management,
resource allocation, and network planning to name a few. Radio maps
are constructed from measurements collected by spectrum sensors
distributed across space. Since radio  maps are complicated
functions of the spatial coordinates due to the nature of
electromagnetic wave propagation, model-free approaches are strongly
motivated. Nevertheless, all existing schemes \firstrev{for radio
occupancy map estimation} rely on interpolation
algorithms unable to learn from \secondrev{experience}. In contrast, this paper proposes
a novel approach in which the spatial structure of propagation
phenomena such as shadowing is learned beforehand from a data set with
measurements in other environments. Relative to existing schemes, a
significantly smaller number of measurements is therefore required to
estimate a map with a prescribed accuracy. As an additional novelty,
this is also the first work to estimate radio \firstrev{occupancy} maps using deep neural
networks. Specifically, a \secondrev{fully convolutional} deep completion autoencoder architecture is
developed to effectively exploit the manifold structure of this class of maps.
\end{abstract}
\section{Introduction}
\label{sec:introduction}
\cmt{Motivation}
\begin{myitemize}
\myitem\cmt{cartography overview}Spectrum cartography comprises a
collection of techniques to construct radio maps, which provide
channel metrics such as received signal power, interference power,
power spectral density (PSD), electromagnetic absorption, or channel
gain across a geographic area; see
e.g.~\cite{alayafeki2008cartography,bazerque2010sparsity,jayawickrama2013compressive,
  yilmaz2013radio}.  \myitem\cmt{motivating applications}%
\begin{myitemize}%
\myitem\cmt{source localization}Besides applications related to
localization~\cite{bazerque2010sparsity,huang2019charting} and radio
tomography~\cite{patwari2008correlated,romero2018blind},
\myitem\cmt{communications}%
\begin{myitemize}%
\myitem\cmt{general}radio maps find a myriad of applications in
wireless communications such as network planning, interference
coordination, power control, spectrum management, resource allocation,
handoff procedure design, dynamic spectrum access, and cognitive
radio; see e.g.~\cite{grimoud2010rem,dallanese2011powercontrol,romero2017spectrummaps}.
\myitem\cmt{UAVs}More recently, radio maps have been widely recognized as an
enabling technology for UAV communications because they allow
autonomous UAVs to account for communication constraints when planning
a mission; see e.g.~\cite{zhang2019path,romero2019noncooperative,bulut2018disconnectivity,chen2017map}.
\end{myitemize}
\end{myitemize}%
\end{myitemize}

\cmt{Literature review}%
\begin{myitemize}%
\myitem\cmt{radio propagation prediction}\firstrev{The problem of predicting  the received
signal strength at a given location or channel gain between a pair
of locations, \secondrev{termed \emph{radio propagation prediction,}} has been initially tackled through modeling}. 
\begin{myitemize}
\myitem\cmt{Model-based}\firstrev{This
includes empirical models~\cite[Ch. 2]{goldsmith}, the dominant path model\cite{wahl2005dominant}, and the more
sophisticated ray-tracing
algorithms (see e.g.~\cite{schaubach1992raytracing}) to name a few. \thirdrev{Due
to the inherent limitations of modeling,
\myitem\cmt{Data-driven}data-driven alternatives that learn from   channel
measurements have been recently developed~\cite{parera2020tiltdependent,imai2019radiopredictioncnn,
saito2019twosteppathloss, hayashi2020studyradiopropagation,
iwasaki2020transferbasedpower,levie2019radiounet,levie2020pathlossprediction}}}.\myitem\cmt{limitations}
\begin{myitemize}%
\myitem\cmt{known tx position, power, number}\firstrev{The main limitation of
these approaches 
is that the number, locations, and (generally) power
of all transmitters need to be known.}
\myitem\cmt{interference}\secondrev{Besides, these data-driven approaches are
sensitive to interference since they assume that the measurements
contain only power received from a single  transmitter with known
power and location.
        }\myitem\cmt{computational complexity}\firstrev{Furthermore, \secondrev{most of these schemes for radio propagation prediction suffer from a large computational
        complexity when predictions must be provided at a large set of
        locations.}
        }
\end{myitemize}
\end{myitemize}

\myitem\cmt{radio map estimation}\firstrev{To address these limitations,
approaches for \emph{radio (occupancy)\footnote{\secondrev{The term
``occupancy'' is brought here from the cognitive radio literature,
where it is used to refer to the aggregate contribution of all
transmitters at the same time, frequency, and spatial location; see
e.g. \cite{agarwal2018spectrum}.}} map estimation} target the
aggregate power from all active transmitters and predict power or
channel gain across the area of interest} \secondrev{without knowledge
of the number, locations, and power of the
transmitters.} \secondrev{This is a critical  benefit especially in
scenarios with a large number of transmitters, as in the uplink of a
cellular network or in device-to-device communications, since this
information is typically unreliable and costly to maintain.}
\begin{myitemize}%
\myitem\cmt{power maps}\firstrev{Most } \secondrev{approaches for radio occupancy map estimation} build upon  some interpolation
algorithm. For example,
\begin{myitemize}%
\myitem\cmt{kriging}power maps have been constructed through
kriging~\cite{alayafeki2008cartography,boccolini2012wireless, agarwal2018spectrum, romero2020surveying},
\myitem\cmt{compress. sensing, dict. lear., matrix
  completion}compressive sensing~\cite{jayawickrama2013compressive}, dictionary learning~\cite{kim2011link,kim2013dictionary}, matrix
completion~\cite{ding2016cellular},
\myitem\cmt{bayes. models, radial basis functions, and kernel
  methods}Bayesian models~\cite{huang2015cooperative}, radial basis functions~\cite{hamid2017non, zha2018spectrum}, and kernel
methods~\cite{teganya2019locationfree}.
\end{myitemize}%
\myitem\cmt{PSD maps}%
\begin{myitemize}%
\myitem\cmt{sparsity}PSD map estimators have been
developed using sparse learning~\cite{bazerque2010sparsity},
 \myitem\cmt{kernel methods}thin-plate spline
regression~\cite{bazerque2011splines}, kernel-based
learning~\cite{bazerque2013basispursuit, romero2017spectrummaps},
 \myitem\cmt{tensor compl, decomp}and tensor completion~\cite{tang2016spectrum,zhang2019spectrum}.
\end{myitemize}
\myitem\cmt{other metrics \ra channel gain maps}Related approaches have been adopted  in~\cite{kim2011cooperative,lee2016lowrank,
  romero2018blind,lee2019bayesian}   to propose
channel-gain map estimators.
\myitem\cmt{Limitations}%
\begin{myitemize}%
  \myitem\cmt{learn from experience}Unfortunately, none of \secondrev{these approaches for radio occupancy map estimation 
  }  
 can learn from \secondrev{experience}, which suggests that their estimation
performance can be significantly improved along this direction.
\footnote{The conference version of this
  work~\cite{teganya2020autoencoders} presents the core ideas
  here. Relative to~\cite{teganya2020autoencoders}, the present paper
  contains improved neural network architectures (\secondrev{namely
    fully convolutional networks}), a methodology for PSD estimation
  with basis expansion models, and extensive empirical validation and
  comparison with existing algorithms through a realistic data
  set. \fourthrev{The work in~\cite{han2020power, levie2019radiounet,
      levie2020pathlossprediction}, based also on deep learning estimators, was conducted in parallel
    to~\cite{teganya2020autoencoders}.  Whereas~\cite{han2020power}
    proposes a generative adversarial network scheme for radio occupancy map
    estimation, \cite{levie2019radiounet,levie2020pathlossprediction}
    target radio propagation prediction.}}

\end{myitemize}
\end{myitemize}
\end{myitemize}

\cmt{Contributions}
\begin{myitemize}
\myitem\cmt{paradigm}\firstrev{To this end,  this work puts forth a data-driven paradigm for radio occupancy map estimation.  }The idea is to
learn the spatial structure of relevant propagation phenomena such as
shadowing, reflection, and diffraction using a data set of past
measurements in different environments. Intuitively, learning how
these phenomena evolve across space can significantly reduce the
number of measurements required to achieve a given estimation
accuracy. This is a critical aspect since the time interval in which
measurements are collected needs to be sufficiently short relative to
the temporal variations of the target map in real-world scenarios
(coherence time).
\myitem\cmt{DNN algorithm}\firstrev{The main  contribution of this work is
to build upon this idea to develop} a PSD map estimation algorithm
based on a deep neural network. To cope with the variable number of
measurements, a tensor completion task is formulated based on a
spatial discretization of the area of interest and addressed by means
of a \emph{completion network} with an encoder-decoder
architecture, \thirdrev{which belongs to a broad family of networks with
well-documented merits in other wireless communications
contexts~\cite{felix2018autoencoder}}. This structure is motivated by
the observation that radio maps lie close to a low-dimensional
manifold embedded in a high-dimensional space.
\myitem\cmt{generalizes well to different distrib.}
 Extensive experiments using a
realistic data set obtained with Remcom's Wireless InSite simulator
reveal that the proposed algorithm markedly outperforms
state-of-the-art radio map estimators 
\firstrev{and exhibits strong out-of-sample generalization capabilities.}
\if\singlecol1
\myitem\cmt{code and data}\firstrev{This
data set is available along with the code at~\texttt{https://github.com/yvestegnya2/deep-autoencoders-cartography}.}
\else
\myitem\cmt{code and data}\firstrev{This
data set is available along with the code at~\texttt{https://github.com/yvestegnya2/deep-\newline autoencoders-cartography}.}
\fi
\end{myitemize}

\cmt{emphasize novelty}The novelty of this work is twofold:
    \begin{myitemize}%
      \myitem\cmt{data-driven}(i) it is the first to propose
      data-driven \firstrev{radio \secondrev{occupancy} map estimation}  \secondrev{by learning from experience};   \myitem\cmt{deep
        learning}(ii) it is the first  to propose a deep
      learning algorithm for radio \secondrev{occupancy} map estimation.\footnote{
        \fourthrev{Note that previous and concurrent works in \emph{radio propagation prediction} have also applied deep learning techniques.}}
    \end{myitemize}


\cmt{Paper structure}The rest of this paper is organized as follows.
Sec.~\ref{sec:model} formulates the problem of PSD
map estimation. Sec.~\ref{sec:propmethod} introduces the aforementioned
data-driven radio map estimation paradigm and proposes a deep neural
network architecture based on  completion autoencoders. Simulations, \firstrev{
conclusions, and a discussion} are  provided in Secs.~\ref{sec:numtest}
and~\ref{sec:conclusion}.

\cmt{notation}\emph{Notation:}
    \begin{myitemize}
      \myitem\cmt{cardinality}$|\mathcal{A}|$ denotes the cardinality
      of set $\mathcal{A}$. \myitem\cmt{scalars, vectors}Bold uppercase
(lowercase) letters denote matrices or tensors (column vectors),  \myitem\cmt{i-th entry of vector, i,j-th entry of
        matrix/tensor} $[\bm a]_i$ is the $i$-th entry of
      vector $\bm a$, $[\bm A]_{i,j}$ is the $(i,j)$-th entry of
      matrix $\bm A$, and $[\bm B]_{i,j,k}$ is the $(i,j,k)$-th
      entry of tensor $\bm B$.  \myitem\cmt{transpose}Finally, $\bm
      A^\top$ is the transpose of matrix $\bm A$.
    \end{myitemize}

\section{PSD Map Estimation Problem}
\label{sec:model}

\cmt{Overview}This section formulates the problem of \emph{PSD map}
estimation. The problem where \emph{power maps} are estimated can
be recovered as a special case of  \emph{PSD map} estimation in a
single frequency.

\cmt{signal propagation overview}
\begin{myitemize}
  \myitem\cmt{sources}Consider $\sourcenum$ transmitters, or sources, located in a
  geographic region of interest $\mathcal{X}\subset\rfield^2$ and
  operating in a certain frequency band. Let
  $\transmitpsd_{\sourceind}(f)$ denote the transmit PSD of the $\sourceind$-th source and
  \myitem\cmt{channel}let $H_{\sourceind}(\bm x, f)$ represent the
  frequency response of the channel between the $\sourceind$-th source
  and a receiver with an isotropic antenna at location $\bm x\in
  \mathcal{X}$.
  \myitem\cmt{temporally stationary}Both
  $\transmitpsd_{\sourceind}(f)$ and $H_{\sourceind}(\bm x, f)$ are
  assumed to remain constant over time; see Remark~\ref{rem:locations}.

\myitem\cmt{measurement model}
      \begin{myitemize}
        \myitem\cmt{test location}If the $\sourcenum$ signals are
        uncorrelated, the PSD at $\bm x \in \mathcal{X}$ is 
      \begin{align}
    \label{eq:receivedpsd}
     \truepsd(\bm x, f)=\textstyle\sum_{\sourceind=1}^{\sourcenum} \transmitpsd_{\sourceind}(f)\vert H_{\sourceind}(\bm x, f) \vert^2 + \upsilon(\bm x, f),
    \end{align}
    where $\upsilon(\bm x, f)$ models thermal noise, background
    radiation noise, and interference from remote sources.
    \myitem\cmt{measurement locations}A certain number of devices with
    sensing capabilities, e.g. user terminals
    in a cellular network, collect PSD measurements $\{ \measpsd (\bm
    x_n, f)\}_{n=1}^N$ at $N$ locations $\{\bm x_n\}_{n=1}^N \subset
    \mathcal{X}$ and at a finite set of frequencies $f\in \mathcal{F}$; see
    also Remark~\ref{rem:locations}.  These frequency measurements can be
    obtained using e.g. periodograms or spectral analysis methods such
    as the Bartlett or Welch method~\cite{stoica2005}.
      \end{myitemize}
      
      \myitem\cmt{fusion center}These measurements are sent to a
      fusion center, which may be e.g. a base station, a mobile user, or a
      cloud server, depending on the application. 
    \end{myitemize}%
\cmt{spectrum cartography problem}%
\begin{myitemize}%
      \myitem\cmt{given}Given $\{(\bm x_n, \measpsd(\bm
      x_n, f) ),~n=1,\ldots,N,~f\in\mathcal{F}\}$,
      \myitem\cmt{requested}the problem that the fusion center needs
      to solve is to find an estimate $\estimatepsd(\bm x, f)$ of
      $\truepsd(\bm x, f)$ at every location $\bm x \in
      \mathcal{X}$ and frequency $f\in \mathcal{F}$.
      \myitem\cmt{terminology}Function $\truepsd(\bm x, f)$ is
      typically referred to as the \emph{true map}, whereas
      $\estimatepsd(\bm x, f)$ is the \emph{map estimate}. An
      algorithm that produces $\estimatepsd(\bm x, f)$ is termed
      \emph{map estimator}.
\end{myitemize}%

    \cmt{challenge}%
    \begin{myitemize}%
      \myitem\cmt{minimize error}A natural error metric is the energy
      $\sum_f\int_\mathcal{X} | \truepsd(\bm x, f) -
      \estimatepsd(\bm x, f) |^2d\bm x$. One can quantify the
      performance of a map estimator in terms of the expectation of
      this error for a given $N$ or, equivalently, in terms of the
      minimum $N$ required to guarantee that the expected 
      error is below a prescribed bound.
          \end{myitemize}

\cmt{fading averaged out}
  \begin{myremark}%
\label{rem:fading}The  channel $H_{\sourceind}(\bm x, f)$  is usually
decomposed into path loss, shadowing, and fast fading
components. Whereas path loss and shadowing typically vary in a scale
of meters, fast fading changes in a scale comparable to the
wavelength. Since contemporary wireless communication systems utilize
wavelengths in the order of millimeters or centimeters, estimating
this fast fading component would require knowing the sensor locations
$\{\bm x_n\}_n$ with an accuracy in the order of millimeters, which is
not possible e.g.  with current \emph{global navigation satellite
  systems} (GNSSs).  Thus, it is customary to assume that the effects
of fast fading have been averaged out and, hence, $H_{\sourceind}(\bm
x, f)$ captures only path loss and shadowing. This is especially
well-motivated in scenarios where sensors acquire measurements while
moving.
  \end{myremark}

      \cmt{coherence time}
      \begin{myremark}
        \label{rem:locations}
         $\transmitpsd_{\sourceind}(f)$ and $H_{\sourceind}(\bm x, f)$
        can be assumed constant over time so long as the measurements
        are collected within an interval of shorter length than the
        channel coherence time and time scale of changes in
        $\transmitpsd_{\sourceind}(f)$. The latter is highly dependent
        on the specific application. For example, one expects that
        significant variations in DVB-T bands occur in a scale of
        several months, whereas $\transmitpsd_{\sourceind}(f)$ in LTE
        bands may change in a scale of milliseconds. In any case, a sensor
        that moves may collect multiple measurements over this
        interval where $\transmitpsd_{\sourceind}(f)$ and
        $H_{\sourceind}(\bm x, f)$ remain approximately constant,
        which could render the number of measurements significantly
        larger than the number of sensors.
      \end{myremark}

\section{Data-Driven Radio Map Estimation}
    \label{sec:propmethod}

    \cmt{data-driven psd map estimation problem}
    \begin{myitemize}
      \myitem\cmt{motivation \ra existing approaches}All existing \secondrev{occupancy} map
      estimators rely on interpolation algorithms that do not learn
      from \secondrev{experience}. However, it seems natural that an algorithm can be
      trained to learn how to solve the problem in
      Sec.~\ref{sec:model} using a record of past measurements,
      possibly in other geographic regions. 
      \myitem\cmt{learning problem formulation}Specifically, besides
    \begin{myitemize}%
      \myitem\cmt{Given}%
      \begin{myitemize}%
        \myitem\cmt{test} $\dataset\define\{(\bm x_n, 
        \measpsd(\bm x_n, f) ),\bm x_n \in
        \mathcal{X},~f\in\mathcal{F},~n=1,\ldots,N\}$,
        \myitem\cmt{training}a number of measurement records of the
        form $\dataset_t\define\{(\bm x_{nt}, 
        \measpsd_t(\bm x_{nt}, f) ),~\bm x_{nt}\in
        \mathcal{X}_t,~f\in\mathcal{F},~n=1,\ldots,N_t\}$, $t=1,
        \ldots, T$, may be available, where $\dataset_t$ contains
        $N_t$ measurements collected in the geographic area $
        \mathcal{X}_t$; see Sec.~\ref{sec:realworld}.
      \end{myitemize}
      \myitem\cmt{Requested}With this additional data, a better performance is
      expected when estimating $\truepsd(\bm x, f)$.
    \end{myitemize}

    \end{myitemize}
    
    \cmt{section overview}The rest of this section  develops deep learning
    estimators  that address this data-aided  formulation. To this end,
    Sec.~\ref{sec:tensorcompletion} starts by reformulating the
    problem at hand as a tensor completion task amenable to
    application of deep neural networks. Subsequently,
    Sec.~\ref{sec:missing} addresses unique aspects of tensor/matrix
    completion via deep learning. Sec.~\ref{sec:exploitstruc} discusses how to exploit structure in the frequency domain. Finally, Secs.~\ref{sec:convautoenc} and~\ref{sec:realworld} respectively describe how to learn the spatial structure of propagation phenomena via the
    notion of \emph{completion autoencoders} and how these networks can be trained in real-world scenarios.

    \subsection{Map Estimation as a Tensor Completion Task}
    \label{sec:tensorcompletion}
    \cmt{motivation}
    \begin{myitemize}
      \myitem\cmt{challenge}Observe that  $N$ and $N_t$ depend on the
    number and movement of the sensors relative to  the time-scale of
    temporal  variations in $\truepsd(\bm x,f)$ and $\truepsd_t(\bm x,f)$, respectively;
    cf. Remark~\ref{rem:locations}.
    \myitem\cmt{Possible approaches}%
        \begin{myitemize}\myitem\cmt{many vs one}          
        \begin{myitemize}\myitem\cmt{separate estimators}In principle, one could think of using a
    separate map estimator for each possible value of $N$. Each
    estimator could be relatively simple since it would always take
    the same number of inputs. However, such an approach would be
    highly inefficient in terms of memory, computation, and prone to
    erratic behavior since each estimator would have different
    parameters or be trained with a different data set.
    \myitem\cmt{one estimator}Thus, it is more practical to rely on a
    single estimator that can accommodate an arbitrary number of measurements.
        \end{myitemize}
        \end{myitemize}
        
        \begin{myitemize}%
          \myitem\cmt{deep learning}Given their well-documented merits
          in a number of tasks, deep neural networks constitute a
          sensible framework to develop radio map estimators. However,
          regular feedforward neural networks cannot directly
          accommodate inputs of variable size.
          \begin{myitemize}%
            \myitem\cmt{tensor completion}To bypass this difficulty,
            the approach pursued here relies on a spatial
            discretization amenable to application of feedforward
            architectures~\cite[Ch. 6]{goodfellow2016deep}. Similar discretizations have been  applied in~\cite{romero2018blind,hamilton2014modeling,ding2016cellular,tang2016spectrum}.
    \end{myitemize}
        \end{myitemize}
        \end{myitemize}

    \begin{figure}[t]
      \centering
      \if\singlecol1
      \includegraphics[scale=0.3]{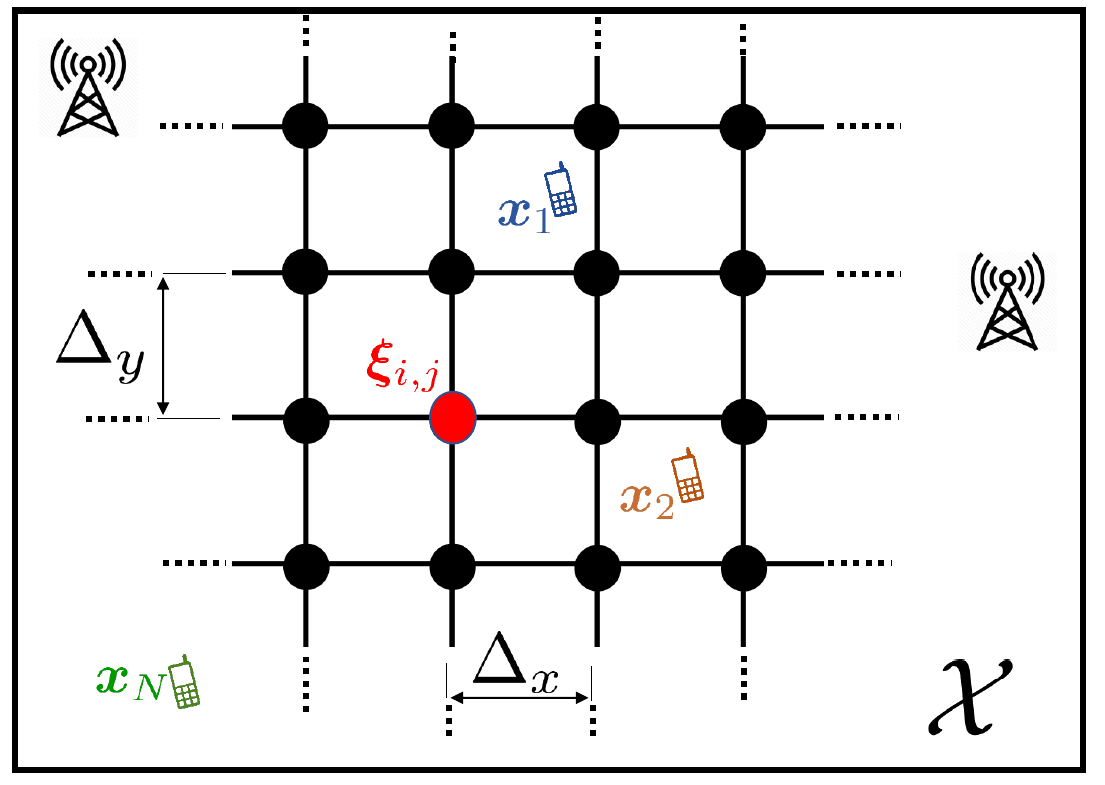}
      \else
      \includegraphics[scale=0.3]{figures/grid_fig.pdf}
      \fi
    \caption{Model setup and area discretization.} 
    \label{f:setupwithgrid}
    \end{figure}

    \cmt{spatial discretization}To introduce the appropriate notation,
    this discretization is briefly outlined for $\dataset$; the
    extension to $\dataset_t$ follows the same
    lines.  \begin{myitemize} \myitem\cmt{2D grid}Define an
    $N_y \times N_x$ rectangular grid over $\mathcal{X}$, as depicted
    in Fig.~\ref{f:setupwithgrid}. This grid comprises points
    $\gridpoint_{i,j}$ evenly spaced by $\Delta_x$ and $\Delta_y$
    along the $x$- and $y$-axes respectively, that is, the $(i,j)$-th
    grid point is given by $\gridpoint_{i,j}:=\left
    [j\Delta_x,~i\Delta_y\right]^\top$, with $~i=1,\ldots,N_y
    ,~j=1,\ldots,N_x$.  \myitem\cmt{assign meas. loc to grid pts}For
    future usage, define $\mathcal{A}_{i,j} \subset \{1,\ldots,N\}$ as
    the set containing the indices of the measurement locations
    assigned to the $(i,j)$-th grid point by the criterion of minimum
    distance,\footnote{\firstrev{Note that multiple measurements acquired
    by the same sensor may be assigned to the same grid point if its
    speed is small in terms of $\Delta_x$, $\Delta_y$, and the time
    between measurements.}} i.e., $n\in \mathcal{A}_{i,j}$ iff
    $||\gridpoint_{i,j}-\bm x_n||\leq ||\gridpoint_{i',j'}-\bm
    x_n||~\forall i',j'$.  \end{myitemize}

      \cmt{True map}This grid induces a discretization of $ \truepsd
      (\bm x, f) $ along the $\bm x$ variable. One can therefore
      collect the true PSD values at the grid points in matrix $\bm
      \truepsd (f) \in \rfield^{N_y \times N_x}$, $f\in \mathcal{F}$,
      whose $(i,j)$-th entry is given by $\left[ \bm \truepsd (f)
        \right]_{i,j}= \truepsd(\gridpoint_{i,j}, f)$. By letting
      $\mathcal{F}=\{f_1,\ldots,f_{N_f}\}$, it is also possible to
      concatenate these matrices to form the tensor $\bm \truepsd\in
      \rfield^{ N_y\times N_x\times N_f}$, where $\left[ \bm \truepsd
        \right]_{i,j,n_f}= \truepsd(\gridpoint_{i,j}, f_{n_f})$,
      $n_f=1,\ldots,N_f$. For short, the term \emph{true map} will
      either refer to $ \truepsd(\bm x, f)$ or $\bm \truepsd$.

      \cmt{measurements}Similarly, one can collect the measurements in
      a tensor of the same dimensions. 
    \begin{myitemize}
          \myitem\cmt{motivate aggregation}Informally, if the grid is
          sufficiently fine ($\Delta_x$ and $\Delta_y$ are
          sufficiently small), it holds that $\bm x_n \approx
          \gridpoint_{i,j}$ $\forall n \in \mathcal{A}_{i,j}$ and,
          correspondingly, $ \truepsd(\bm x_n, f) \approx
          \truepsd(\gridpoint_{i,j}, f)$ $\forall n \in
          \mathcal{A}_{i,j}$. It follows that,
$
       \truepsd(\gridpoint_{i,j}, f)\approx({1}/{\vert \mathcal{A}_{i,j} \vert)}\sum_{n \in \mathcal{A}_{i,j} }\truepsd(\bm x_n, f)
       $
          whenever $\vert \mathcal{A}_{i,j} \vert >0 $.       
          \myitem\cmt{aggregation}Therefore, it makes sense to aggregate the measurements
      assigned to $\gridpoint_{i,j}$ as\footnote{For simplicity, the
        notation implicitly assumes that $\bm x_n\neq
        \gridpoint_{i,j}$ $\forall n,i,j$, but this is not a requirement.}
$
        \measpsd(\gridpoint_{i,j}, f)\define ({1}/{\vert
         \mathcal{A}_{i,j} \vert})\sum_{n \in \mathcal{A}_{i,j} }
        \measpsd(\bm x_n, f)$.
~\firstrev{Observe that aggregation of this form could alleviate the
effects of small scale fading in the channel; see Remark~\ref{rem:fading}.} 
       \myitem\cmt{misses} Conversely, when $|\mathcal{A}_{i,j}|=0$,
       there are no measurements associated with
       $\gridpoint_{i,j}$, in which case one says that there is a
       \emph{miss} at        $\gridpoint_{i,j}$.
       \myitem\cmt{sampled map matrix}Upon letting $\Omega \subset \{1,\ldots,N_y\}\times \{1,\ldots, N_x\}$ be 
such that $(i,j) \in \Omega$ iff $\vert \mathcal{A}_{i,j} \vert > 0$, all  aggregated
       measurements $ \measpsd(\gridpoint_{i,j}, f)$ can be
       collected into  $\measpsdmat (f) \in \rfield^{N_y
  \times N_x}$, defined as $[\measpsdmat (f)
         ]_{i,j}=       
       \measpsd(\gridpoint_{i,j}, f)$ if $
                  (i,j) \in \Omega$ and $[ \tilde {\bm \truepsd}
         (f) ]_{i,j}=0$ otherwise. 
\myitem\cmt{filling}Note that misses  have been
filled with zeroes, but other values could have been used. 

    
      \myitem\cmt{error sources}When $(i,j)\in \Omega$,
      the values of $ [ \tilde {\bm \truepsd} (f) ]_{i,j}$
      and $ [ {\bm \truepsd} (f) ]_{i,j}$ differ due to the
      error introduced by the spatial discretization as well as due to
      the measurement error incurred when measuring
      $\truepsd(\bm x_n,f),~n\in \mathcal{A}_{i,j}$. The latter is
      caused mainly by the finite time devoted by sensors to take 
      measurements, their movement, localization errors, and possible
      variations of $\truepsd(\bm x_n,f)$  over time.

      \myitem\cmt{tensor}As before, the matrices
      $\measpsdmat(f),~f=1,\ldots,N_f$ can be concatenated to form
      $\measpsdmat \in \rfield^{ N_y\times N_x\times N_f}$, where
      $[\measpsdmat ]_{i,j,n_f} = [\measpsdmat (f_{n_f})
      ]_{i,j}$. For short, this tensor will be referred to as the
      \emph{sampled map}.
    \end{myitemize}
    \cmt{Problem reformulation}With this notation, the cartography
    problem stated in Sec.~\ref{sec:model} will be approximated as
    \begin{myitemize}%
      \myitem\cmt{req.}estimating $\bm \truepsd$
      \myitem\cmt{given}given $\Omega$ and $\measpsdmat$,
\end{myitemize}



    \subsection{Completion Networks for Radio Map Estimation}
    \label{sec:missing}
    
\cmt{overview}The data in the problem formulation at the end of
Sec.~\ref{sec:tensorcompletion} cannot be  handled by
plain feedforward neural networks since they cannot directly accommodate
input misses and set-valued inputs like $\Omega$. This section
explores how to bypass this difficulty. 

\cmt{DL refresh}Before that, a swift refresh on deep learning is
in order.  A feedforward deep neural network is a function $\nnfun$
that can be
\begin{myitemize}%
  \myitem\cmt{layers}expressed as the composition $
  \nnfun(\auxinmat)= \layerfun\layerparnot{\layernum}{\bm
    w_{\layernum}}(\layerfun\layerparnot{\layernum-1}{\bm
    w_{\layernum-1}}(\ldots\layerfun\layerparnot{1}{\bm
    w_{1}}(\auxinmat))) $ of \emph{layer} functions
  $\layerfun\layerparnot{\layerind}{\bm w_\layerind}$, where
  $\auxinmat$ is the input. Although there is no commonly agreed
  definition of layer function, it is typically formed by
  concatenating simple scalar-valued functions termed \emph{neurons}
  that implement a linear aggregation followed by a non-linear function
  known as activation~\cite{goodfellow2016deep}.
  \myitem\cmt{neural}The term \emph{neuron} stems from the resemblance
  between these functions and certain simple functional models for
  biological neurons.  \myitem\cmt{deep}Similarly, there is no general
  agreement on which values of $\layernum$ qualify for $\nnfun$ to be
  regarded a \emph{deep} neural network, but in practice $\layernum$
  may range from tens to thousands.  \myitem\cmt{parameters}With
  vector $\bm w_\layerind$ containing the parameters, or weights, of the
  $\layerind$-th layer, the parameters of the entire network can be
  stacked as $\bm w \define [\bm w_1\transpose, \ldots,\bm
    w_{\layernum}\transpose]\transpose \in \rfield^{N_w}$.
  \myitem\cmt{training}These parameters are \emph{learned} using a
  \emph{training set} in a process termed \emph{training}.

    \end{myitemize}%

\cmt{Outlook}%
\begin{myitemize}%
  \myitem\cmt{Training in our application \ra outlook}The rest of this
  section designs  $\layerfun\layerparnot{1}{\bm w_{1}}$ to cope with
  missing data, whereas  Secs.~\ref{sec:exploitstruc}
  and~\ref{sec:convautoenc} will address the design of the other layers.  
  \myitem\cmt{training data}Throughout, the \emph{training examples}
  will be represented by $\{( \measpsdmat_t, \Omega_t)\}_{t=1}^{T}$,
  where $\measpsdmat_t$ and $\Omega_t$ are obtained from $\dataset_t$
  by applying the procedure described in
  Sec.~\ref{sec:tensorcompletion}.
\end{myitemize}%

\cmt{Dealing with misses}%
\begin{myitemize}%
  \myitem \cmt{maximize w.r.t. missing
    entries} \begin{myitemize} \myitem\cmt{description}The desired
    estimator should obtain $ \bm \truepsd$ as a function of \firstrev{the
    measurements in} $\measpsdmat$ and $\Omega$. \firstrev{Recall that
    only a subset of the entries of $ \measpsdmat$ contain actual
    measurements, whereas the rest have been filled, e.g. with
    zeros. Unfortunately, it is not directly possible to accommodate
    variable input supports in regular feedforward neural
    networks. However, as a first attempt,}\footnote{\firstrev{One could
    alternatively think of completing the misses through optimization
    variables as in~\cite{fan2017deep}. However, the resulting number of
    variables would grow with the size of the data
    set, thereby rendering such an approach impractical.}}%
\myitem\cmt{limitations \ra complexity \ra prohibitive}    
  \end{myitemize}%
  \myitem\cmt{not minimize, feed zeroes or another filling value.}\firstrev{one
  could think of directly
  feeding $\measpsdmat$ to the neural network and training by only
  fitting the entries containing measurements as }
      \begin{myitemize}%
      \myitem\cmt{description}
  \begin{myitemize}%
    \myitem\cmt{training}
    \begin{align}  \label{eq:matricompw}
      \underset{\bm w}{ \text{minimize}}\quad
      &\textstyle\frac{1}{T}\sum_{t=1}^T
      \left \Vert \mathcal{P}_{\Omega_t}\left( \measpsdmat_t- p_{\bm w}(\measpsdmat_t
      )\right)  \right \Vert_F^2,
    \end{align}
where $||\bm A||_F^2\define
\sum_{i,j,n_f}[\bm A]_{i,j,n_f}^2$ is the squared Frobenius norm of
tensor $\bm A$ and $\mathcal{P}_{\Omega}(\bm A)$ is defined by
$\left[ \mathcal{P}_{\Omega}(\bm A) \right]_{i,j,n_f}= [\bm
A]_{i,j,n_f}$ if $(i,j) \in \Omega$ and
$\left[ \mathcal{P}_{\Omega}(\bm A) \right]_{i,j,n_f}= 0$
otherwise.
    \myitem\cmt{testing}After \eqref{eq:matricompw} is solved, $\measpsdmat$ could be completed just by evaluating $p_{\bm w}(\measpsdmat)$.
 \myitem\cmt{complexity control}\firstrev{Observe that if the
family of candidate functions $\{p_{\bm w}: \bm w \in \rfield^{N_w}\}$
contains the identity map $p_{\bm w}(\measpsdmat)=\measpsdmat$, then
the minimum of \eqref{eq:matricompw} would be attained for such a
function. This would clearly render the estimator useless. Thus, one needs to
introduce some form of complexity control~\cite{cherkassky2007} e.g. by regularization or by limiting the family  $\{p_{\bm w}: \bm
w \in \rfield^{N_w}\}$. }\myitem\cmt{autoencoder structure}\firstrev{This work pursues the
    second strategy by means of an encoder-decoder architecture, as
    detailed in Sec.~\ref{sec:convautoenc}.}
    \end{myitemize}%
  \end{myitemize}%
    \begin{myitemize}%
    \myitem    \begin{myitemize}%
      \myitem\cmt{strengths}

    \end{myitemize}%

      \end{myitemize}%
      
    \myitem\cmt{limitation}Because the aforementioned estimator does not account for $\Omega$, poor
    performance is expected since  the network cannot distinguish missing entries
    from  measurements close to the filling value.
    \end{myitemize}%
\myitem\cmt{natural units}In the application at hand, one could
circumvent this limitation by expressing the entries of $\measpsdmat$ in natural power units (e.g. Watt) and filling the misses
with a negative number such as -1. Unfortunately, the usage of
finite-precision arithmetic would introduce large errors in the map
estimates and is problematic in our experience. For this reason,
expressing $\measpsdmat$ in logarithmic units such as dBm is
preferable. However, in that case, filling misses with negative
numbers would not solve the aforementioned difficulty since 
logarithmic units are not confined to take non-negative values.
      \myitem\cmt{complement input with a mask \ra add a channel}Hence, a
  preferable alternative is to complement the input map with a
  binary mask that indicates which entries are observed, as proposed
  in the image inpainting literature~\cite{iizuka2017consistent}.
  \begin{myitemize}%
    \myitem\cmt{mask}Specifically, a mask $\bm \mask_{\Omega} \in
    \{0,1\}^{N_y \times N_x}$ can be used to represent  $\Omega$  by setting $[\bm \mask_{\Omega}]_{i,j}=1$ if $ (i,j)
    \in \Omega$ and $[\bm \mask_{\Omega}]_{i,j}= 0$ otherwise.
\myitem\cmt{mask as a channel}To simplify notation, let  $\augmeaspsdmat\in
\rfield^{N_y\times N_x \times N_f+1}$ denote a tensor obtained by concatenating $\measpsdmat$ and
$\bm \mask_{\Omega}$ along the third dimension.
\myitem\cmt{training}The neural network can therefore be trained as
\begin{align}  \label{eq:matricompmask}
      \underset{\bm w}{ \text{minimize}}\quad
      &\textstyle \frac{1}{T}\sum_{t=1}^T
      \left \Vert \mathcal{P}_{\Omega_t}\left( \measpsdmat_t- p_{\bm w}(\augmeaspsdmat_t
      )\right)  \right \Vert_F^2,
    \end{align}
    \myitem\cmt{testing}and, afterwards, a tensor $\measpsdmat$ can
    be completed just by evaluating $p_{\bm w}(\augmeaspsdmat )$.
    \myitem\cmt{strengths}Then, this scheme is simple to train,
    inexpensive to test, and exploits information about the location
    of the misses. 
\end{myitemize}%

\begin{figure}[t]
\begin{center}
\if\singlecol1
\includegraphics[width=0.6\columnwidth]{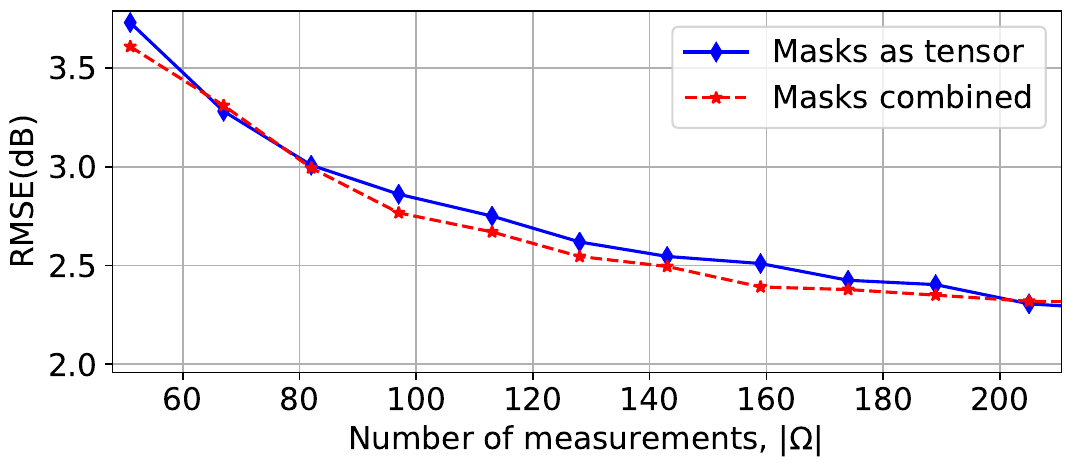}
\else
\includegraphics[width=1\columnwidth]{figures/rmse_masks.pdf}
\fi
\caption{\secondrev{Map estimation RMSE as a function of the number of
measurements when the building and measurement masks are separately
provided (\emph{masks as tensor}) or combined into a single mask
(\emph{masks combined}). The training and testing maps were obtained
from the Wireless InSite data set; see Sec.~\ref{sec:numtest}.}}
\label{f:rmse_2maskCong}
\end{center} 
\end{figure}

\cmt{meta info}
\begin{myremark}
\label{rem:meta}The introduction of a binary mask to indicate the
  sensor locations suggests an approach \firstrev{along the lines
  of~\cite{imai2019radiopredictioncnn, iwasaki2020transferbasedpower}
  to accommodate additional side information that may assist in map
  estimation}.
\begin{myitemize}%
  \myitem\cmt{meta map}For example, one can append an additional mask
  $\bm \metamap \in \rfield^{N_y \times N_x}$ where
  $[\bm \metamap]_{i,j}$ indicates \firstrev{e.g.} the height of
  obstacles such as buildings at $\gridpoint_{i,j}$ or the kind of
  propagation terrain (e.g. urban, suburban, etc) where
  $\gridpoint_{i,j}$ lies.  \begin{myitemize} \myitem\cmt{metamap as
  other channels}In this case, tensor
  $\augmeaspsdmat \in \rfield^{N_y\times N_x \times N_f+1}$ can be
  replaced with its augmented version of size $N_y\times
  N_x \times(N_f+N_m)$ obtained by concatenating $N_m-1$ such masks to
  $\measpsdmat$ and $\bm \mask_{\Omega}$.  \myitem\cmt{mask and
  metamap combined}Another (possibly complementary) approach is to
  combine multiple masks into a single matrix. For example, suppose
  that all measurements are taken outdoors and let
  $\building \subset \{1,\ldots,N_y\}\times \{1,\ldots, N_x\}$ be such
  that $(i,j) \in \building$ iff $\gridpoint_{i,j}$ is inside a
  building. Then, the information in $\building$ and $\Omega$ can be
  combined into $\bm \mask_{\Omega, \building} \in \{0,1,
  -1\}^{N_y \times N_x}$, where
  $[\bm \mask_{\Omega, \building}]_{i,j}=1$ if $ (i,j) \in \Omega$,
  $[\bm \mask_{\Omega, \building}]_{i,j}= -1$ if
  $(i,j) \in \building$, and $[\bm \mask_{\Omega, \building}]_{i,j}=
  0$ otherwise. Masks of this kind can be similarly concatenated to
  $\measpsdmat$ to \firstrev{form}
  $\augmeaspsdmat$. \secondrev{Combining masks has the benefit of
  reducing the number of parameters to train without sacrificing
  performance, as illustrated in Fig.~\ref{f:rmse_2maskCong}; the
  details about the network and simulation setup can be found in
  Sec.~\ref{sec:numtest}. 
  } The rest of the paper
  will \firstrev{use} $\augmeaspsdmat$ to refer to the result of
  concatenating $\measpsdmat$ with the available masks.

\end{myitemize}%
\end{myitemize}
\end{myremark}

\begin{myremark}
  \label{rem:dbfitting}
The proposed deep learning framework offers an additional
advantage: the tensors in the objective functions throughout
(e.g. \eqref{eq:matricompw}, \eqref{eq:matricompmask}) can be
expressed in dB units. This is not possible in most existing
approaches such as \firstrev{~\cite{bazerque2010sparsity, kim2011link,bazerque2011splines,
  kim2011cooperative, jayawickrama2013compressive, kim2013dictionary,
  ding2016cellular,tang2016spectrum, lee2016lowrank, hamid2017non,
  romero2017spectrummaps, zha2018spectrum,
  romero2018blind,teganya2019locationfree, zhang2019spectrum}}, which rely on convex solvers. Consequently,
existing algorithms focus on fitting large power values and
neglect errors at those locations with low power values. Given its
greater practical significance, it will be assumed throughout that all
tensors are expressed in dB units before evaluating the Frobenius norms. 

\end{myremark}

\subsection{Exploiting Structure in the Frequency Domain}
\label{sec:exploitstruc}

\cmt{overview}In practice, different degrees of prior information may
be available when estimating a PSD map. Sec.~\ref{sec:noprior} will
address the scenario in which no such information is available, whereas
Sec.~\ref{sec:completionbem} will develop an output layer that
exploits a common form of prior information available in real-world
applications.

\subsubsection{No Prior Information}
\label{sec:noprior}
\cmt{Alleviating Ill-Posedness\ra frequency separation}It will be
first argued that the plain training approach in
\eqref{eq:matricompmask} is likely to be ill-posed in practical
scenarios when the network does not enforce or exploit any structure
in the frequency domain.
\begin{myitemize}%
  \myitem\cmt{issue}To see this, suppose that the number of
  frequencies $N_f$ in $\mathcal{F}$ is significant, e.g. 512 or 1024
  as would typically occur in practice, and consider a fully connected first layer
  $\layerfun\layerparnot{1}{\bm w_1}$ with $N_N$ neurons. Its total
  number of parameters becomes $(N_yN_x(N_f+N_m)+1)N_N$ plus possibly
  additional parameters of the activation functions. Other layers will
  experience the same issue to different extents.  Since $T$ must be
  comparable to the number of parameters to train a network
  effectively, a large $N_f$ would drastically limit the number of
  layers or neurons that can be used for a given~$T$.
  \myitem\cmt{separation}
  \begin{myitemize}%
    
\myitem\cmt{separation}\firstrev{In absence of further prior information, one
  possibility to reduce the number of parameters is} to separate the
  problem across frequencies by noting that propagation effects at
  similar frequencies are expected to be similar. \firstrev{To understand
  this, it is instructive to note that the maximum difference in the
  free-space path loss between the lowest and the highest frequencies
  is $\Delta P_\text{Rx}= 20\text{log}_{10}((f_c+BW/2)
  /(f_c-BW/2))$ where $f_c$ and $BW$ are respectively the carrier
  frequency and bandwidth. For example, when $f_c=1400$ MHz and
  $BW=10$ MHz, $\Delta P_\text{Rx}\approx 0.06$ dB. 
  Although fading effects may differ due to constructive/destructive
  interference, one can claim in narrowband scenarios that
  propagation approximately affects all frequencies in the same
  fashion.}  Building upon this principle, $\nnfun$ can operate
  separately at each frequency $f$. This means that training can be
  accomplished through \myitem\cmt{train}
\begin{align} \label{eq:fsep}
  \underset{\bm w}{ \text{minimize}}~
  &\frac{1}{T N_f}\sum_{t=1}^T\sum_{f\in \mathcal{F}}
  \left \Vert \mathcal{P}_{\Omega_t}\left( \measpsdmat_t(f)- p_{\bm w}(\augmeaspsdmat_t(f)
  )\right)  \right \Vert_F^2, 
\end{align}
    where the input $\augmeaspsdmat_t(f)\in \rfield^{N_y\times N_x \times (1 +
      N_m)}$ is formed by concatenating $\measpsdmat_t(f)$ and $N_m$
    masks; see Remark~\ref{rem:meta}.

    \myitem\cmt{strengths}
    \begin{myitemize}%
      \myitem\cmt{variables}Observe that the number of variables is
      roughly reduced by a factor of $N_f$, whereas \myitem\cmt{data
        examples} the ``effective'' number of training examples has been
      multiplied by $N_f$; cf. number of summands in
      \eqref{eq:fsep}. This is a drastic improvement especially
    for moderate values of $N_f$.  Thus, such a frequency separation allows an
       increase in the number of neurons per layer or (typically more
      useful~\cite[Ch. 5]{goodfellow2016deep}) the total number of layers for a
      given $T$. Although such a network
      would not exploit structure across the frequency domain, the
      fact that it would be better trained  is likely
      to counteract this limitation in many setups. 
    \end{myitemize}%

          \end{myitemize}%

      \end{myitemize}%

\subsubsection{Output Layers for Parametric PSD Expansions}
    \label{sec:completionbem} 
    \cmt{motivation}%
    \begin{myitemize}%
      \myitem\cmt{standards}Real-world communication systems typically
      adhere to standards that specify transmission masks by means of
      carrier frequencies, channel bandwidth, roll-off factors, number
      of OFDM subcarriers, guard bands, location and power of pilot
      subcarriers, and so on. It seems, therefore, reasonable to
      capitalize on such prior information for radio map estimation by
      means of a basis expansion model in the frequency domain like
      the one in~\cite{vazquez2011guardbands, romero2013wideband, romero2017spectrummaps}.
      \myitem\cmt{approximation}Even when the frequency form of the
      transmit PSD is unknown, a basis expansion model is also
      motivated due to its capacity to approximate any PSD to some extent;
      e.g.~\cite{bazerque2011splines,bazerque2010sparsity}.
\end{myitemize}
    \if\singlecol1
     \begin{figure}[t!]
    \centering
    \includegraphics[width=0.75\columnwidth]{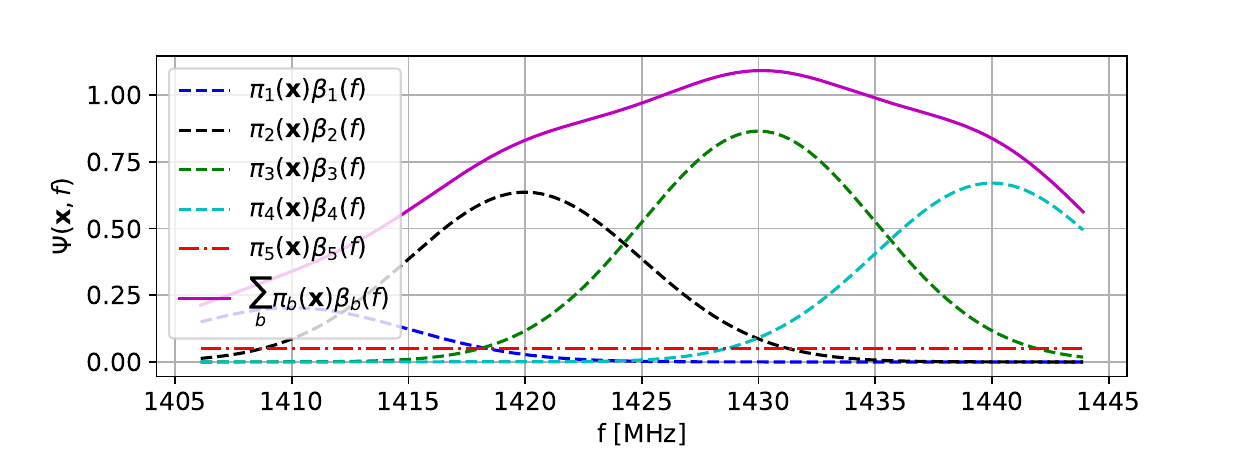}
    \caption{PSD $\truepsd (\bm x, f)$ at location $\bm x$ using a basis expansion model with Gaussian functions.}
    \label{f:freq_bases}
  \end{figure}
    \else
       \begin{figure}[t!]
    \centering
    \includegraphics[width=1\columnwidth]{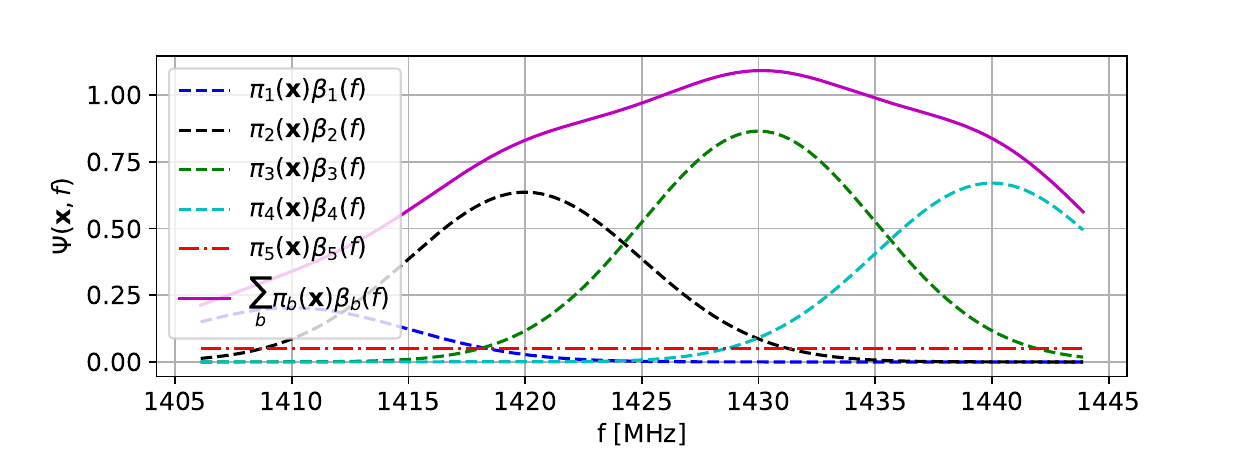}
    \caption{PSD $\truepsd (\bm x, f)$ at location $\bm x$ using a basis expansion model with Gaussian functions.}
    \label{f:freq_bases}
  \end{figure}
  \fi

\cmt{BEM}Under a basis expansion model,
\begin{myitemize}%
  \myitem\cmt{sources}the transmit PSD of each source is expressed as
  \begin{align}
    \label{eq:txpsdbem}
\transmitpsd_{\sourceind}(f)=\textstyle\sum_{\baseind=1}^{\basenum-1}
\truecoeffs_{\sourceind \baseind} \bases_{\baseind}(f),
    \end{align}
 where
$\truecoeffs_{\sourceind \baseind}$ denotes the expansion coefficients
and $\{ \bases_{\baseind}(f)\}_{\baseind=1}^{\basenum-1}$ is a
collection of $\basenum-1$ given  basis functions such as raised-cosine
or Gaussian functions. Without loss of generality, the basis functions
are normalized so that
$\int_{-\infty}^{\infty}\bases_{\baseind}(f)df=1$. In this way, if
$\bases_{\baseind}(f)$ is the PSD of the $\baseind$-th channel, as
possibly specified by a standard,
then $\truecoeffs_{\sourceind \baseind} $ denotes the power
transmitted by the $\sourceind$-th source in the $\baseind$-th channel.
\myitem\cmt{measurement model}%
\begin{myitemize}%
\myitem\cmt{test location}Substituting \eqref{eq:txpsdbem} into
\eqref{eq:receivedpsd}, the PSD at $\bm x \in \mathcal{X}$ reads as
    \begin{align*}
      \truepsd(\bm x, f)&
  =\sum_{\sourceind=1}^{\sourcenum}\sum_{\baseind=1}^{\basenum-1} \truecoeffs_{\sourceind \baseind} \bases_{\baseind}(f)\vert H_{\sourceind}(\bm x, f) \vert^2 + \upsilon(\bm x, f).
   \end{align*}
   \myitem\cmt{freq. flat}Now assume that $\vert H_{\sourceind}(\bm x,
   f) \vert^2$ remains approximately constant over the support of each
   basis function, i.e., $\vert H_{\sourceind}(\bm x, f)
   \vert^2\approx \vert H_{\sourceind \baseind}(\bm x) \vert^2$ for
   all $f$ in the support of $\bases_{\baseind}(f)$. This is a
   reasonable assumption for narrowband $\bases_{\baseind}(f)$; if it
   does not hold, one can always split $\bases_{\baseind}(f)$ into
   multiple basis functions with a smaller frequency support until the
   assumption holds.  Then, the PSD at $\bm x$ can be written as
   \begin{align}
     \label{eq:rxpsdbem}
 \truepsd(\bm x, f)=\sum_{\baseind=1}^{\basenum-1} \truecoeffs_{\baseind}(\bm x) \bases_{\baseind}(f) + \upsilon(\bm x, f),
 \end{align} 
 where $\truecoeffs_{\baseind}(\bm
 x)\define\textstyle\sum_{\sourceind=1}^{\sourcenum}
 \truecoeffs_{\sourceind \baseind} \vert H_{\sourceind \baseind}(\bm
 x) \vert^2$. If $\bases_{\baseind}(f)$ models the transmit PSD of the
 $\baseind$-th channel, then $\truecoeffs_{\baseind}(\bm x)$
 corresponds to the power of the $\baseind$-th channel at $\bm x$.

 \myitem\cmt{noise in the bem model}Observe that the noise PSD
 $\upsilon(\bm x, f)$ can be similarly expressed in terms of a basis
 expansion. To simplify the exposition, suppose that $\upsilon(\bm x,
 f)$ is expanded with a single term as $\upsilon(\bm x, f)\approx
 \truecoeffs_{\basenum}(\bm x)\bases_\basenum(f)$, which in turn
 implies that \eqref{eq:rxpsdbem} becomes
 \begin{align}
 \truepsd(\bm x, f)=\sum_{\baseind=1}^{\basenum} \truecoeffs_{\baseind}(\bm x) \bases_{\baseind}(f).
 \end{align} 
Fig.~\ref{f:freq_bases} illustrates this expansion for $\basenum=5$
when $\{\bases_\baseind(f)\}_{\baseind=1}^4$ are Gaussian radial basis
functions and $\bases_5(f)$ is set to be constant to model the PSD of white
noise.
Note that the adopted basis expansion
furthermore allows estimation of the noise power $
\truecoeffs_{\basenum}(\bm x)$ at every location. \firstrev{This is of
 special interest in  applications such as cognitive
radio~\cite{tandra2008snrwalls}.
}
\end{myitemize}%
\end{myitemize}%

\cmt{output layer}With the above expansion, 
\begin{myitemize}%
\myitem\cmt{true map}the tensor $\bm \truepsd\in \rfield^{ N_y\times
  N_x\times N_f}$ introduced in Sec.~\ref{sec:tensorcompletion} can be
expressed  as $\left[ \bm \truepsd
  \right]_{i,j,n_f}=\sum_{\baseind=1}^{\basenum} \truecoeffsmat_{i,j,
  \baseind} \bases_{\baseind}(f_{n_f})$, where $\truecoeffsmat \in
\rfield^{ N_y\times N_x\times \basenum}$ contains the 
coefficients $\left[ \truecoeffsmat \right]_{i,j,\baseind} =
\truecoeffs_{\baseind}(\gridpoint_{i,j})$.
  \myitem\cmt{split}In a deep neural network, this structure
  can be naturally enforced by setting all but the last layer to
  obtain an estimate $\estimatecoeffsmat$ of $\truecoeffsmat$ and the
  last layer to produce $\estimatepsdmat$. Specifically, the neural network can be expressed schematically as:
    \begin{alignat*}{3}
\begin{aligned}  
&\mathcal{L} &  \overset{\nnbemfun}{\xrightarrow{\hspace*{0.7cm}}}
& \quad  \rfield^{ N_y\times N_x\times \basenum} & \overset{\layerfun\layerparnot{\layernum}{}}{\xrightarrow{\hspace*{0.7cm}}}&   \quad  \rfield^{ N_y\times N_x \times N_f}
\\
&\augmeaspsdmat &  \xrightarrow{\hspace*{0.7cm}}  & \quad \estimatecoeffsmat & \xrightarrow{\hspace*{0.7cm}}& \quad \estimatepsdmat,
\end{aligned}
    \end{alignat*}
    \cmt{where}where
    \begin{myitemize}
      \myitem\cmt{}$\mathcal{L} \subset \rfield^{N_y\times N_x \times(
        N_f + N_m)}$ is the input space,
      \myitem\cmt{} function $\nnbemfun(\augmeaspsdmat) \define \layerfun\layerparnot{\layernum-1}{\bm
    w_{\layernum-1}}(\ldots\layerfun\layerparnot{1}{\bm
    w_{1}}(\augmeaspsdmat ))$ groups the first
      $\layernum -1$ layers,
      \myitem\cmt{}and  $\layerfun\layerparnot{\layernum}{}$ denotes the last layer.
      \myitem\cmt{}With this notation,  $\estimatecoeffsmat=
      \nnbemfun(\bm \augmeaspsdmat)$
      and $ \estimatepsdmat = \layerfun\layerparnot{\layernum}{}(\estimatecoeffsmat) \in  \rfield^{ N_y\times N_x
      \times N_f}$, where $[\estimatepsdmat]_{i,j,n_f}=
    \sum_{\baseind=1}^{\basenum} \estimatecoeffsmat_{i,j, \baseind}
    \bases_{\baseind}(f_{n_f})$.
      \end{myitemize}
    \myitem\cmt{Observations}
        \begin{myitemize}%
          \myitem\cmt{no parameters last layer}Observe that, as
          reflected by the notation, the last layer
          $\layerfun\layerparnot{\layernum}{}$ does not involve
          trainable parameters.  \myitem\cmt{total net
            parameters}Furthermore, notice that the number of neurons
          in the last trainable layer has been reduced from
          $N_yN_xN_f$ to $N_yN_x\basenum$. 

\firstrev{To sum up, this section presented two possibilities to reduce the number of parameters
of the network, which can improve estimation performance for a given
      training size. Whereas the approach in Sec.~\ref{sec:noprior} is
      more suitable for narrowband channels, the approach in
      Sec.~\ref{sec:completionbem} lends itself to wideband
      communications, where propagation effects may significantly
      differ across frequencies.}  \end{myitemize}


    
\end{myitemize}%

\subsection{Deep Completion Autoencoders}
\label{sec:convautoenc}
\cmt{Overview}The previous section addressed design aspects pertaining
to the map structure in the frequency domain.  In contrast, this
section deals with structure across space. In particular, a deep
neural network architecture based on \secondrev{fully}~\emph{convolutional
  autoencoders}~\cite{ribeiro2018study} will be developed.

\cmt{autoencoder}
  \begin{myitemize}%
    \myitem\cmt{conventional AE}A (conventional) autoencoder~\cite[Ch. 12]{goodfellow2016deep} is a neural
    network $\nnfun $ that can be expressed as the composition of a
    function  $\encfun$ termed \emph{encoder} and a function 
     $\decfun$ called \emph{decoder}, i.e., 
    $\nnfun(\auxinmat)=\decfun(\encfun(\auxinmat))~\forall\auxinmat$.
    The output of the encoder $\code\define\encfun(\auxinmat)\in
    \rfield^{\latentnum}$ is referred to as the \emph{code} or vector
    of \emph{latent variables} and is of a typically much lower
    dimension than the input $\auxinmat$. An autoencoder is trained so
    that $\decfun(\encfun(\auxinmat))\approx \auxinmat ~\forall
    \auxinmat$, which forces the encoder to compress the information
    in $\auxinmat$ into the $\latentnum$ variables in
    $\code$. 

    \myitem\cmt{Completion AE}
  \begin{myitemize}%
    \myitem\cmt{description}A \emph{completion} autoencoder adheres to
    the same principles as conventional autoencoders except for the
    fact that the encoder must determine the latent variables from a
    subset of the entries of the input. If a  mask is used, then
    $
\auxinmat \approx \decfun(\encfun(\mathcal{P}_{\Omega}(\auxinmat),\bm
\mask_\Omega))~\forall\auxinmat
$
 if the  sampling set $\Omega$ preserves sufficient
information for reconstruction -- if $\Omega$ does not satisfy this
requirement, then reconstructing $\auxinmat$ is impossible regardless
of the  technique used.  \myitem\cmt{this application}In the
application at hand and with the notation introduced in previous
sections, the above expression becomes
$\mathcal{P}_{\Omega}(\measpsdmat) \approx \mathcal{P}_{\Omega}( \decfun(\encfun( \augmeaspsdmat )))$.

  \end{myitemize}%

  \end{myitemize}%
\begin{figure}[t!]
\centering
\includegraphics[width=1\columnwidth]{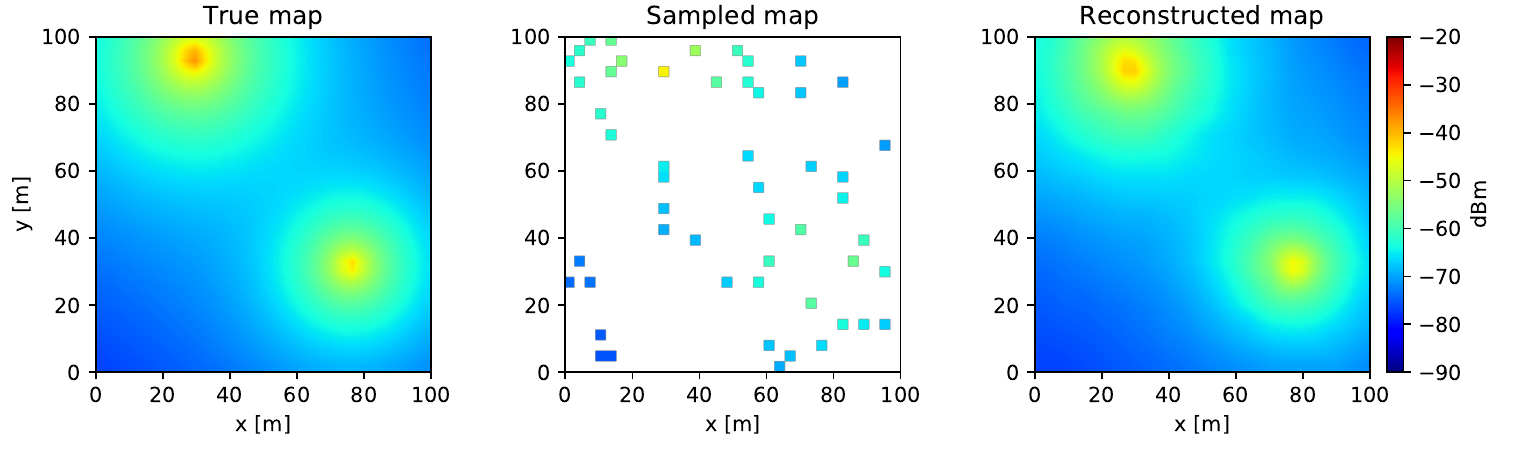}
\caption{Estimation with $N_\lambda=4$ latent
  variables: (left) true map, (middle) sampled map portraying 
   grid points $ \{\gridpoint_{i,j} \}$ with $\vert
  \mathcal{A}_{i,j}\vert > 0 $, and (right) map estimate. }
\label{f:motivatingex}
\end{figure}
\cmt{latent variables}
\begin{myitemize}%
  \myitem\cmt{requirement for autoencoders}As indicated earlier,
  autoencoders are useful only when most of the information in the
  input can be condensed in $\latentnum$ variables, i.e., when the
  possible inputs lie close to a manifold of dimension $\latentnum$.
  \myitem\cmt{example}To see that this is indeed the case in radio map
  estimation, an illustrating toy example is presented next.
    \begin{myitemize}%
    \myitem\cmt{description}Consider two sources transmitting with a
    different but fixed power at arbitrary positions in $\mathcal{X}$
    and suppose that propagation occurs in free space. All possible
    spectrum maps in this setup can therefore be uniquely identified
    by $\latentnum=4$ quantities, namely the x and y coordinates of
    the two sources.  \myitem\cmt{figure}Fig.~\ref{f:motivatingex}
    illustrates this effect, where the left panel of
    Fig.~\ref{f:motivatingex} depicts a true map $\bm \truepsd$ and
    the right panel shows its estimate using the proposed completion
    autoencoder when $\latentnum=4$. Although the details about the
    network and simulation setup are deferred to 
    Sec.~\ref{sec:numtest},  one can already notice at this point
    the quality of
    the estimate, which clearly supports the aforementioned manifold
    hypothesis.    \myitem\cmt{shadowing}In a
    real-world scenario, there may be more than two sources, their
    transmit power may not always be the same, and there are shadowing
    effects, which means that $\latentnum\geq 4$ will be generally
    required.
    \end{myitemize}%
\end{myitemize}%

\begin{figure}[ht!]
\centering
      \if\singlecol1
      \includegraphics[scale=0.5]{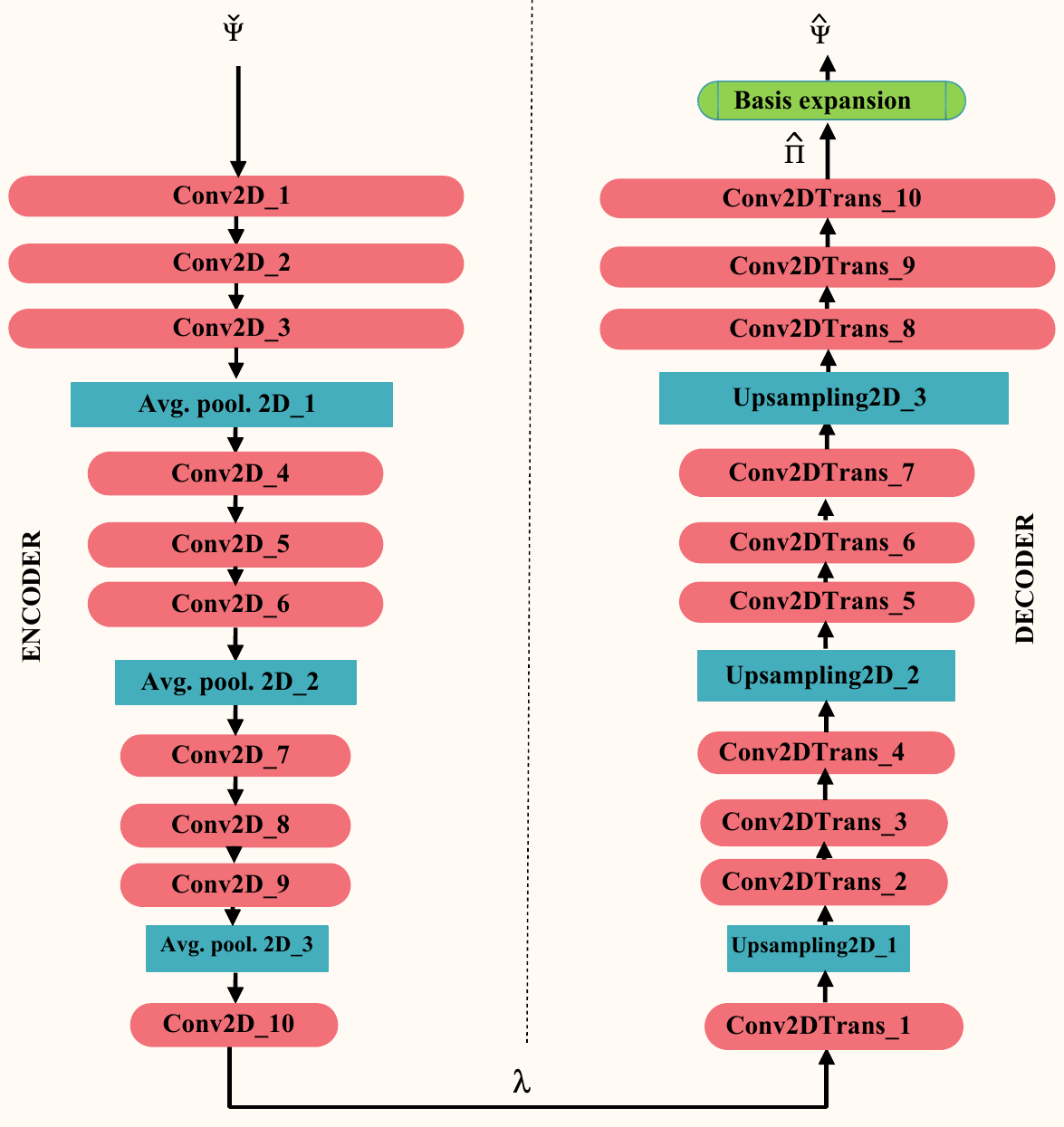}
      \else
      \includegraphics[width=1\columnwidth]{figures/autoencoder_model.pdf}
      \fi

\caption{Autoencoder architecture.} 
\label{f:modelarch}
\end{figure}

\cmt{Architecture}The rest of this section will describe the main aspects of the architecture developed in this work
and summarized in Fig.~\ref{f:modelarch}. The main design decisions are supported here by arguments and intuition. Empirical support is provided in Sec.~\ref{sec:netdesign}.

\begin{myitemize}%
  \myitem\cmt{encoder}The encoder mainly comprises convolutional and
  pooling layers. 
  \begin{myitemize}%
    \myitem \cmt{convolutional layers}%
    \begin{myitemize}%
      \myitem\cmt{motivation}%
      The motivation for convolutional
      layers is three-fold:
      \begin{myitemize}%
        \myitem\cmt{motivation}
\cmt{reduce no. params. }(i) relative to fully connected
          layers, they severely reduce the number of parameters to train and, consequently,
          the amount of data required.
\cmt{shift-invariance \ra convolutional NN}Despite this
          drastic reduction, (ii) convolutional layers are still capable of
          exploiting the spatial structure of maps and (iii) they result in
          shift-invariant transfer functions, a desirable property in the
          application at hand since moving the sources in a certain direction
          must be corresponded by  the same movement in the map estimate. 
      \end{myitemize}%
      \myitem\cmt{description}Recall that a convolutional layer with
      input $\auxlayerin$ and output
      $\auxlayerout$ linearly combines  2D convolutions as
      \begin{align}
        \label{eq:2dconvxx}
        [\auxlayerout]_{i, j,\outchind} =\sum_{\inchind=1}^\inchnum \sum_{u=-k}^{k}\sum_{v=-k }^{k} [\bm F_\outchind]_{u, v,\inchind}[\auxlayerin]_{ i -u, j -v,\inchind},
      \end{align}
      where
    \begin{myitemize}%
      \myitem{}the $\outchind$-th \emph{kernel} $\bm F_\outchind$ is of size $2k+1 \times 2k+1\times \inchnum$. 
    \end{myitemize}%
    Layer indices were omitted in order not to
    overload notation. 
      \myitem\cmt{activation \ra PReLUs}The adopted activation functions are parametric
      \emph{leaky rectified linear units} (PReLUs)~\cite{he2015delving}, whose leaky parameter is also trained; see also Sec.~\ref{sec:netdesign}.
    \end{myitemize}%
        \myitem\cmt{pooling}
    \begin{myitemize}%
\myitem\cmt{avgpool}\emph{Average pooling} layers are used to
down-sample the outputs of convolutional layers, thereby condensing
the information gradually in fewer features while approximately
preserving shift invariance~\cite[Ch. 9]{goodfellow2016deep}.
    \end{myitemize}
%
  \end{myitemize}%

  \myitem\cmt{decoder}As usual in autoencoders, the decoder follows a
  ``reverse'' architecture relative to the
  encoder.  \begin{myitemize} \myitem\cmt{conv2DTr}Specifically, for
  each convolutional layer of the encoder, the decoder has a
  corresponding~\emph{ convolution transpose}
  layer~\cite{dumoulin2016guide}, sometimes called ``deconvolutional''
  layer.  \myitem\cmt{upsampling}Likewise, the pooling layers of the
  encoder are matched with \emph{up-sampling} layers. A simple
  possibility is to implement such an upsampling operation by means of
  bilinear interpolation. \firstrev{Alternatively, one can replace a
  convolution transpose layer followed by an interpolation layer with
  a single convolution layer with a fractional stride~\cite{dumoulin2016guide}}.

  \end{myitemize}%
\end{myitemize}%
\begin{table}[t!]
\normalsize 
\caption{Parameters of the proposed network.}
\label{table:autoencoder_par}
\centering 

 \if\singlecol1
 \renewcommand{\arraystretch}{0.8}
      \begin{tabular}{|p{4cm}|p{10cm}|}
\hline
\multirow{1}{5em}{Layers}& Parameters \\ 
\hline
\multirow{1}{5em}{Conv2D/ Conv2DTranspose} & Kernel size = $3\times3$, stride = 1,
activation = PLReLU,   32 filters  \\ 
\hline
\multirow{1}{5em}{AveragePooling2D} & Pool size = 2, stride = 2  \\ 
\hline
\multirow{1}{5em}{Upsampling2D} & Up-sampling factor = $2$, bilinear interpolation \\  
\hline
\end{tabular}

      \else
     \renewcommand{\arraystretch}{1.25} 
     \begin{tabular}{|p{2.5cm}|p{5cm}|}
       \hline
       \multirow{1}{8em}{ Layers}& Parameters \\ 
       \hline
       \multirow{1}{8em}{Conv2D/ Conv2DTranspose} & Kernel size   = $3\times3$, stride = 1,
         activation = PLReLU,   32 filters  \\ 
\hline
\multirow{1}{8em}{AveragePooling2D} & Pool size = 2, stride = 2  \\ 
\hline
\multirow{1}{8em}{Upsampling2D} & Up-sampling factor = $2$, bilinear interpolation \\  
\hline
\end{tabular}

      \fi

\end{table}

\cmt{fully convolutional}Observe that the proposed network, summarized in
Fig.~\ref{f:modelarch} and Table \ref{table:autoencoder_par},  is \emph{fully
  convolutional}, which means that there are no fully connected
layers. This not only leads to a better estimation performance due to
the reduced number of parameters to train
(cf. Sec.~\ref{sec:numtest}), but also enables the possibility of
utilizing the same network with any value of $N_x$ and $N_y$. With fully
connected layers, one would generally require a different network
for each pair $(N_x,N_y)$, which would clearly have negative
implications for training.

\thirdrev{The computational complexity of  the proposed network when
  estimating a radio map is clearly given by the computational
  complexity of a forward pass. The latter is dominated by the
  complexity of the convolutional layers, which, from 
    \eqref{eq:2dconvxx},  can be seen to involve
  $\sum_{\tilde l=1}^{\tilde L}(2k_{\tilde l} +1)^2C_{\text{in},\tilde
  l}C_{\text{out},\tilde l}N_{x,\tilde l}N_{y,\tilde l}$ products in
  total, where $\tilde l$ indexes the $\tilde L$ convolutional layers,
  $k_{\tilde l}$ and $C_{\text{in},\tilde l}$ are such that the kernel
  of the $\tilde l$-th layer is of size $(2 k_{\tilde l} + 1) \times
  (2 k_{\tilde l} + 1) \times C_{\text{in},\tilde l}$, whereas
  $N_{x,\tilde l}$, $ N_{y,\tilde l}$, and $C_{\text{out},\tilde l}$
  are such that the output of the $\tilde l$-th layer is $N_{x,\tilde
  l} \times N_{y,\tilde l} \times C_{\text{out},\tilde l}$. Note that
  these constants are determined by the
  strides, intermediate pooling layers, and the adopted padding
  technique.  }


\subsection{Learning in Real-World Scenarios}
\label{sec:realworld}

\cmt{data-driven}A key novelty in this paper is to obtain map
estimators by learning from data. This section describes how to
construct a suitable training set in the application at hand.
\cmt{training approaches}Specifically, three approaches are discussed:
\subsubsection{Synthetic Training Data}
\label{sec:realworldsynthetic}
\begin{myitemize}%
  \myitem\cmt{idea}Since collecting a large number of training maps
  may be slow or expensive, one can instead generate maps using a
  mathematical model or simulator that captures the structure of the
  propagation phenomena, such as path loss and shadowing; see
  e.g.~\cite{jeruchim2006communication}. Fitting $\nnfun$ to data
  generated by that model would, in principle, yield an estimator that effectively
  exploits this structure.
  \myitem\cmt{approach1 \ra PSD}%
\begin{myitemize}%
  \myitem\cmt{training phase}%
    \begin{myitemize}%
      \myitem\cmt{Data generation}The idea is, therefore, to generate
      $T$ maps $\{ \truepsd_{t}(\bm x, f)\}_{t=1}^T$ together with
      $T$ sampling sets $\{\Omega_t\}_{t=1}^T$. Afterwards, $\{\measpsdmat_t\}_{t=1}^T$ and $\{\augmeaspsdmat_t\}_{t=1}^T$ can be
      formed as described earlier.  \myitem\cmt{denoising}It is
      possible to add artificially generated noise to the synthetic
      measurements in $\measpsdmat_t$ to model the effect of
      measurement error. This would train the network to counteract
      the impact of such error, along the lines of  denoising autoencoders~\cite[Ch. 14]{goodfellow2016deep}.

      \myitem\cmt{obj. func.}The advantage of this
      approach is that one has access to the ground truth, i.e., one
      can use the true maps $\bm \truepsd_{t}$ as
      \emph{targets}. Specifically, the neural network can be trained
      on the data $\{( \augmeaspsdmat_t, \bm
      \truepsd_{t})\}_{t=1}^{T}$ by solving
\begin{align}
\label{eq:objfunction1}
\minimize_{~\bm w}\quad\frac{1}{T}\sum_{t=1}^T
  \left \Vert \bm  \truepsd_{t}-
p_{\bm w}( \augmeaspsdmat_t )\right \Vert_F^2.
\end{align}
    \end{myitemize}%
\myitem\cmt{test phase}If the model or simulator is sufficiently close
to the reality,  completing a real-world  map $\augmeaspsdmat$  as
$\nnfun(\augmeaspsdmat)$ should produce an accurate estimate.
\end{myitemize}%

\myitem\cmt{challenge}
\end{myitemize}%

\subsubsection{Real Training Data}
\label{sec:realworldreal}
\begin{myitemize}%
  \myitem\cmt{approach2}In practice, real maps may be available for
  training. However, in most cases, it will not be possible to collect
  measurements at all grid points within a sufficiently short time interval; see
  Remark~\ref{rem:locations}. Besides, it is not possible to obtain
  the entries of $\bm \truepsd$ but only measurements of it. This
  means that a real training set comprises tensors  $\{ \augmeaspsdmat_t
  ,~t=1,\ldots,T\}$ but not $\bm\truepsd_t$.
  \begin{myitemize}%
    
    \myitem\cmt{train}
    \begin{myitemize}%
      \myitem\cmt{Direct usage}For training, one can plug this data
      directly into \eqref{eq:matricompmask} or
      \eqref{eq:fsep}. However, $\nnfun$ may then learn to fit just
      the observed entries $\{ [\measpsdmat_t(f)]_{i,j},~ (i,j)\in
      \Omega_t \}$, as would happen e.g. when $\nnfun$ is the identity
      mapping.  \myitem\cmt{Sample splitting}To counteract this trend,
      one can adopt a sufficiently small $\latentnum$. The downside is that
       estimation performance may be damaged. To bypass this difficulty,
      the approach proposed here is to use part of the measurements as
      the input and another part as the output (target).
      \begin{myitemize}%
        \myitem\cmt{index sets}Specifically, for each $t$, construct $Q_t$ pairs of (not necessarily disjoint) subsets
        $\Omega_{t,q}^{(I)},\Omega_{t,q}^{(O)}\subset \Omega_t$,
        $q=1,\ldots,Q_t$, e.g by drawing a given number of elements of
        $\Omega_t$ uniformly at random without replacement.
        \myitem\cmt{subsample}Using these subsets, subsample
        $\measpsdmat_t $ to yield $\measpsdmat_{t,q}^{(I)}
        \define \mathcal{P}_{\Omega^{(I)}_{t,q}}(\measpsdmat_{t})$
        and $\measpsdmat_{t,q}^{(O)} \define
        \mathcal{P}_{\Omega^{(O)}_{t,q}}(\measpsdmat_{t})$.
%
%
        \myitem\cmt{obj. func.}With the resulting $\sum_tQ_t$ training instances, one can think of solving
        \if\singlecol1
            \begin{align}\label{eq:objfunction2}
          \minimize_{\bm w} \textstyle\quad\frac{1}{ \sum_tQ_t}\sum_{t=1}^T \sum_{q=1}^{Q_t}           
          \left\|
          \mathcal{P}_{\Omega_{t,q}^{(O)}}\left(
         \tilde {\bm  \truepsd}_{t,q}^{(O)} - 
           p_{\bm w} \left(\augmeaspsdmat_{t,q}^{(I)}\right)
          \right)
          \right\|_F^2,
        \end{align}
        \else
            \begin{align}\label{eq:objfunction2}
          \begin{split}
          \minimize_{\bm w} & \textstyle\quad\frac{1}{ \sum_tQ_t}\sum_{t=1}^T \sum_{q=1}^{Q_t} \\&            
          \left\|
          \mathcal{P}_{\Omega_{t,q}^{(O)}}\left(
         \tilde {\bm  \truepsd}_{t,q}^{(O)} - 
           p_{\bm w} \left(\augmeaspsdmat_{t,q}^{(I)}\right)
          \right)
          \right\|_F^2,
          \end{split}
        \end{align}
        \fi  
      \end{myitemize}%
      where $\augmeaspsdmat_{t,q}^{(I)}
$ is formed by concatenating  $\tilde {\bm
  \truepsd}_{t,q}^{(I)}$  and  $\bm \mask_{\Omega_{t,q}^{(I)}}$.

    \end{myitemize}%
    \myitem\cmt{Limitations}
  \end{myitemize}%
\end{myitemize}%

\subsubsection{Hybrid Training}
\label{sec:hybridtraining}
In practice, one expects to have real data, but only in a limited
amount. It then makes sense to apply the notion of \emph{transfer learning}~\cite[Ch. 15]{goodfellow2016deep} as follows: first, learn an initial parameter vector $\bm
w^*$ by solving~\eqref{eq:objfunction1} with synthetic data. Second,
solve~\eqref{eq:objfunction2} with real data, but using $\bm w^*$ as
initialization for the optimization algorithm. The impact of choosing
this initialization is that the result of
solving~\eqref{eq:objfunction2} in the second step will be generally closer to a
``better'' local optimum than if a random initialization were
adopted. Hence, this approach combines the information of both
synthetic and real data sets.

\section{Numerical Experiments}
\label{sec:numtest}


\begin{myitemize}%
\myitem\cmt{overview}This section validates the proposed framework and
network architecture through numerical experiments.\if\singlecol1\footnote{All code, data sets, and trained networks 
\firstrev{are available at \newline \texttt{https://github.com/yvestegnya2/deep-autoencoders-cartography}.}}
\else\footnote{All code and data sets 
\firstrev{are available at~\newline  \texttt{https://github.com/yvestegnya2/deep-autoencoders- cartography}.}}
\fi

\myitem\cmt{spatial region}\secondrev{Unless stated otherwise,} the region of interest  $\mathcal{X}$ is a square area of side $100$ m, discretized
 into a grid with $N_y=N_x=32$. 
 \myitem\cmt{datasets}Two data sets are constructed as described next.
   \begin{myitemize}%
 \myitem\cmt{gudmundson}First, $T=5\cdot 10^5$ maps are generated
  \begin{myitemize}%
    \myitem\cmt{source}where the two considered transmitters are
    placed uniformly at random in $\mathcal{X}$, have height 1.5 m,
    and transmit with power in each channel drawn uniformly at random
    between $5$ and $11$ dBm.
    \myitem\cmt{channel}%
    \begin{myitemize}%
      \myitem\cmt{pathloss exp}The pathloss exponent is set to 3,
      whereas the \myitem\cmt{gain unit distance}gain at unit distance
      is $-30$ dB.  \myitem\cmt{correlation}The lognormal shadowing
      component adheres to the Gudmundson
      model~\cite{gudmundson1991correlation}, \firstrev{which provides the
        correlation between the shadowing of two links that share an
        endpoint. Specifically, in this paper, } 
      $\expected{H_\sourceind(\bm x_1,f)H_\sourceind(\bm
        x_2,f)}=\sigma_\text{sh}^20.95^{\vert\vert \bm x_1 - \bm x_2
        \vert\vert}$, where $\sigma_\text{sh}^2= 10$ dB$^2$ and
      $\vert\vert \bm x_1 - \bm x_2 \vert\vert$ is the distance
      between $ \bm x_1$ and $\bm x_2$ in meters.
    \end{myitemize}
    \myitem\cmt{sensor locations}Measurement locations are drawn uniformly at
    random without replacement across the grid points. 
    \myitem\cmt{measurements}Each
  measurement $ \measpsd(\bm x_n, f)$ is obtained
  by adding zero-mean Gaussian noise with standard deviation 1 dB to
  $\truepsd(\bm x_n, f)$.
  \end{myitemize}
  
    \myitem\cmt{wireless insite}A second data set of $T=1.25\cdot 10^5$ maps is
    generated using Remcom's Wireless InSite  software
    \begin{myitemize}%
  \myitem\cmt{channel}in the ``urban canyon'' scenario,
  \thirdrev{where a pair of transmitters are deployed per map in the
    downtown of Rosslyn, Virginia. The area is a square of
    approximately 700 m side, out of which $N_x \times N_y$ patches
    are generated by drawing the coordinates of their bottom-left
    corner uniformly at random. } \firstrev{The ray tracing (RT)
    algorithm used there \secondrev{in combination with a 3D map of
      the city} is based on the \emph{shooting and bouncing ray
      method}~\cite{ling1989shooting} (see
    also~\cite[Ch. 8]{espineira2008modeling}) with the maximum number
    of reflections and diffractions set to 6 and 2, respectively}.
  \myitem\cmt{realistic}\firstrev{This data set can be regarded as a
    realistic surrogate of a data set with real measurements due to
    the solid theoretical foundations of RT algorithms on Maxwell's
    equations. Besides, they are extensively employed to construct
    data sets in related works~\cite{imai2019radiopredictioncnn,
      saito2019twosteppathloss,levie2020pathlossprediction,levie2019radiounet,
      iwasaki2020transferbasedpower}.} \myitem\cmt{sensor
    locations}Measurement locations are distributed uniformly at
  random without replacement across the grid points that lie on the
  streets.  \myitem\cmt{map smoothing \ra avg out multipath fading}To
  average out multipath fading present in the generated maps (see
  Remark~\ref{rem:fading}), $ \truepsd(\gridpoint_{i,j}, f)$ is
  replaced with $({1}/{\vert \mathcal{N}_{i,j} \vert)}\sum_{\gridpoint
    \in \mathcal{N}_{i,j} }\truepsd(\gridpoint, f) $, where
  $\mathcal{N}_{i,j}$ contains the $\vert \mathcal{N}_{i,j} \vert=9$
  grid points that lie closest to $\gridpoint_{i,j}$, including
  $\gridpoint_{i,j}$.  \myitem\cmt{mask}A binary mask indicating the
  position of buildings is combined with the sample mask as indicated
  at the end of Remark~\ref{rem:meta}.
      \end{myitemize}
   \end{myitemize}%
   
  \myitem\cmt{Proposed algorithm}The network proposed in
  Sec.~\ref{sec:convautoenc} with code length $\latentnum=64$ is implemented in TensorFlow and trained
  using the Adam solver~\cite{kingma2014adam} with learning rate $5 \cdot 10^{-4}$, \firstrev{batch-size 64, training epochs  100, \secondrev{and the number of measurements $|\Omega_t| ,~t=1,\ldots,T$, in each training map is drawn uniformly at random between 10 and 400}}.
  \myitem\cmt{evaluation metric}Quantitative evaluation will compare
  the root mean square error (RMSE), defined as
  \if\singlecol1
   $\text{RMSE}=
  \sqrt{\mathbb{E}\{ \vert \vert \bm \truepsd-\bm \estimatepsd \vert
      \vert_F ^2 \}/(N_x N_y N_f)}$,     
  \else
   \begin{align}\label{eq:metric}
   \text{RMSE}=
  \sqrt{\frac{\mathbb{E}\{ \vert \vert \bm \truepsd-\bm \estimatepsd \vert
      \vert_F ^2 \}}{N_x N_y N_f}}, 
       \end{align}
  \fi  
  where $\bm \truepsd$ is the true map, $\bm \estimatepsd$ is the map
  estimate, and $\mathbb{E}\lbrace \cdot \rbrace $ denotes expectation
  over noise, and sensor locations.  \firstrev{The RMSE will be estimated
    by averaging over a test data set of $10^3$ maps.}
    \end{myitemize}
      \begin{figure*}[t!]
    \centering\includegraphics[width=\textwidth]{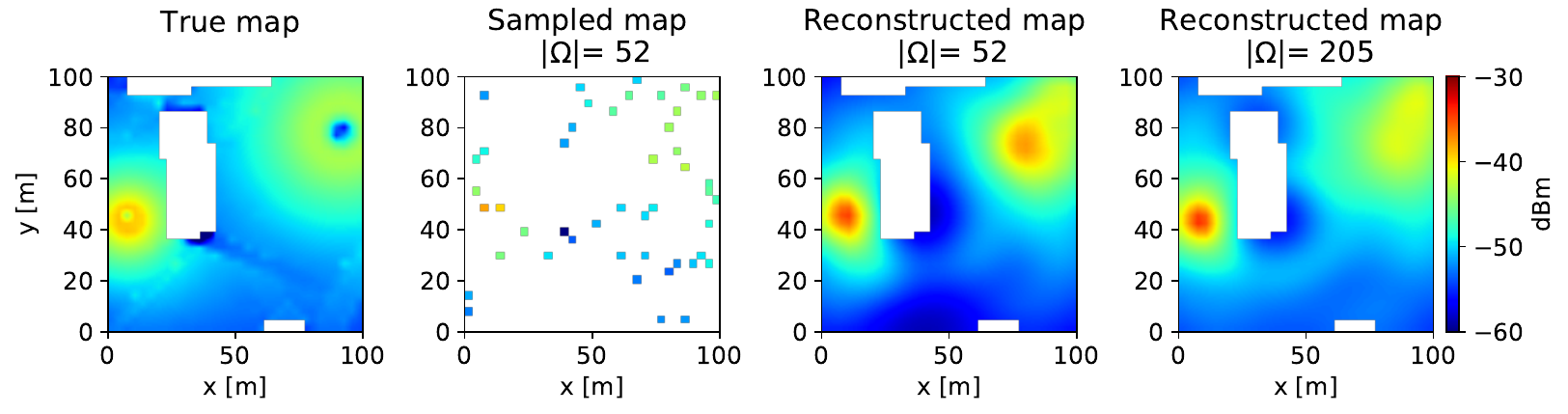}
    \caption{Power map estimate with the proposed neural network.
      (left): true map, (center left): sampled map portraying the locations
      of the grid points $ \{\gridpoint_{i,j} \}$ where $\vert
      \mathcal{A}_{i,j}\vert > 0 $; (center right) and (right):
      map estimates \firstrev{for different numbers of measurements $|\Omega|$}. White areas represent buildings.  }
    \label{fig:reconstructionsample}
  \end{figure*}
 \subsection{Power Map Cartography} 
 \label{sec:experimentspowermap}
 \cmt{overview}To analyze the most fundamental radio map estimation
 aspects, $\mathcal{F}$ is set here to the singleton $\mathcal{F}=\{1400
 \text{ MHz}\}$ and the bandwidth to 5 MHz in both data
 sets.\cmt{noise psd} To better observe the impact of propagation
 phenomena, $\upsilon(\bm x, f)$ is set to 0.
  
\cmt{Competing algorithms}The proposed algorithm is compared against a
representative set of competitors, whose parameters
were adjusted to approximately yield the best performance. This includes:
  \begin{myitemize}%
    \myitem\cmt{alaya-feki}(i) The kriging algorithm
    in~\cite{alayafeki2008cartography} with regularization parameter
    $10^{-5}$ and Gaussian radial basis functions with parameter
    $\sigma_{K} \define 5\sqrt{{\Delta_yN_y\Delta_xN_x}/{\vert
        \Omega\vert}}$, which is approximately 5 times the mean
    distance between two points at which measurements have been
    collected.  \myitem\cmt{multikernel}(ii) The multikernel algorithm
    in~\cite{bazerque2013basispursuit} with regularization parameter
    $10^{-4}$ and \firstrev{two kinds of kernels}: 20 Laplacian kernels that use a parameter uniformly
    spaced between $[0.1\sigma_{K},\sigma_{K}]$ \firstrev{and 20 Gaussian kernels that use different parameters between $0.005$ and $100$ m}. \myitem\cmt{matrix
      completion}(iii) Matrix completion via nuclear norm
    minimization~\cite{ding2016cellular} with regularization
    parameter $10^{-5}$. \myitem\cmt{gaussian proc.}(iv) \firstrev{Gaussian processes for regression~\cite[Ch. 5]{rasmussen2006gaussianprocesses} with regularization parameter
    $3 \cdot 10^{-1}$ and Gaussian radial basis functions.} \myitem\cmt{KNN}As a benchmark, (v) the
    $K$-nearest neighbors (KNN) algorithm with $K=5$ is also shown.
  \end{myitemize}
    \begin{figure}[t!]
    \centering
    \if\singlecol1
      \includegraphics[scale=0.54]{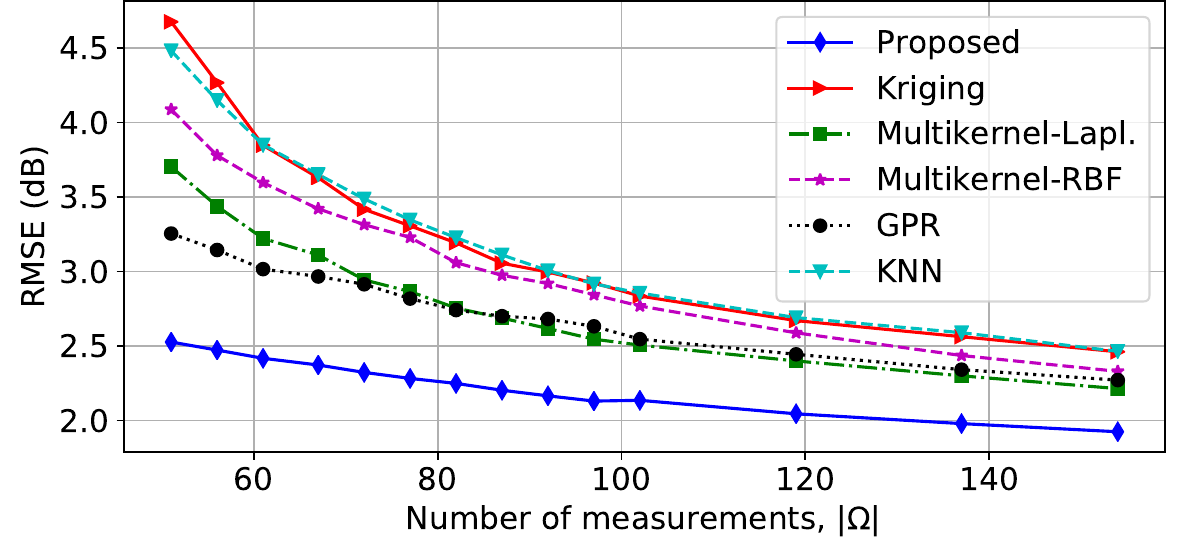}
      \else
      \includegraphics[width=1\columnwidth]{figures/RMSE_vrs_SF_shad_noisy.pdf}
      \fi
    \caption{Comparison with state-of-the-art alternatives. Training and testing maps  drawn from the Gudmundson data set. \secondrev{The proposed network uses a grid with $N_y=N_x=64$, two additional Conv2D and Conv2DTranspose layers, code length $\latentnum=256$, and $T=2.4\cdot 10^6$.} 
    }
    \label{f:comparison}
  \end{figure} 
 \subsubsection{Gudmundson Data Set}
  \label{sec:experimentspowermapgudmund}
\begin{myitemize}%
    \myitem\cmt{training approach}Performance is assessed next  using the training
    approach in Sec.~\ref{sec:realworldsynthetic} with $\{(
    \augmeaspsdmat_t, \bm \truepsd_{t})\}_{t=1}^{T}$ given by the
    Gudmundson data set.
   \end{myitemize}

\cmt{Experiments}
\begin{myitemize}%
  \myitem\cmt{Experiment 1}To analyze estimation of real maps when the
  proposed network is trained over synthetic data, the first
  experiment shows two map estimates when the true (test) map is drawn from
  the Wireless Insite data set. Specifically, the first panel of
  Fig.~\ref{fig:reconstructionsample} depicts the true map, the second
  shows $\measpsdmat$, and the remaining two panels show estimates
  using different numbers of measurements. Observe that with just
  $|\Omega|=52$ measurements, the estimate is already of a high
  quality. Note that details due to diffraction, multipath, and
  antenna directivity are not reconstructed because the Gudmundson
  data set used to train the network does not capture these effects
  and, therefore, the network did not learn them.


  \myitem\cmt{Experiment 2}The second experiment compares the RMSE of the proposed method with that of the competing algorithms.
  \begin{myitemize}%
  \myitem\cmt{Figures}From Fig.~\ref{f:comparison}, the proposed
  scheme performs approximately a \firstrev{30} \% better than the next competing
  alternative. Due to the high RMSE of the matrix completion algorithm
  in~\cite{ding2016cellular} for the adopted range of $|\Omega|$
  in Fig.~\ref{f:comparison}, \firstrev{its RMSE is not shown in
  this figure. 
  }
  \end{myitemize}
  \begin{figure*}[t!]
  \centering
    \if\singlecol1
     \begin{subfigure}{8cm}
     \centering\includegraphics[width=8cm]{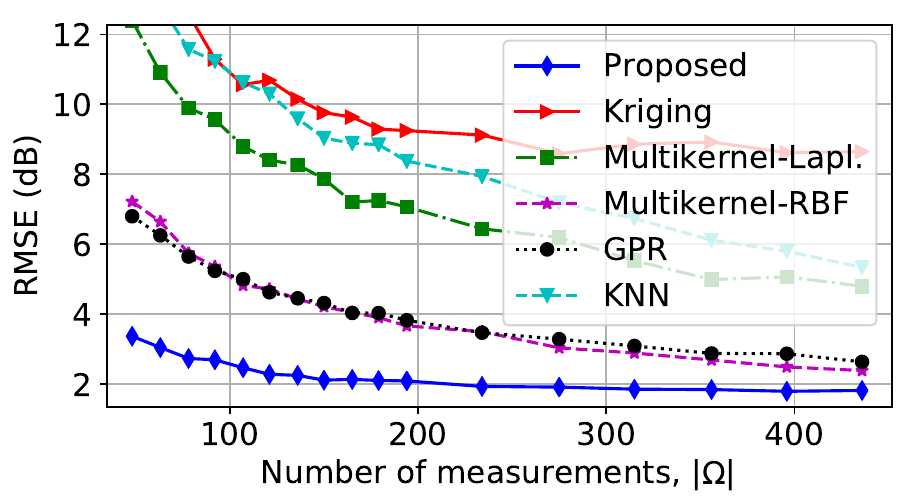}
       \caption{}
        \end{subfigure}
         \begin{subfigure}{8cm}
          \centering\includegraphics[width=8cm]{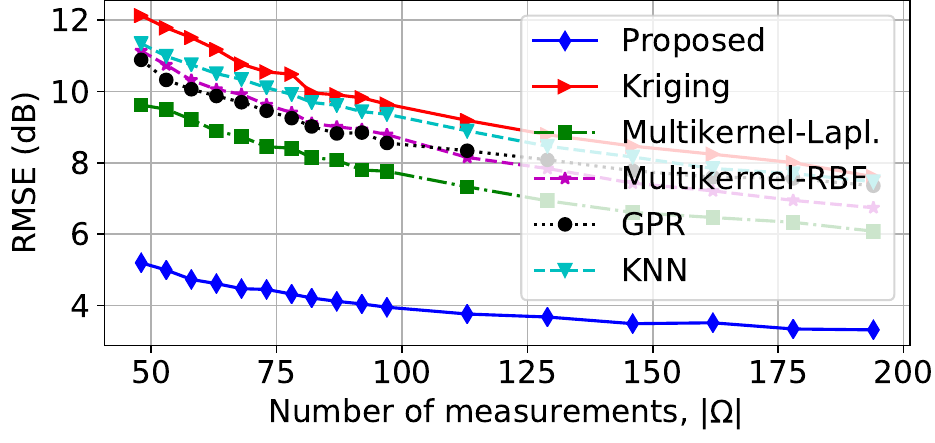}
         \caption{}
        \end{subfigure}
      \else
      \begin{subfigure}{8cm}
      \includegraphics[width=1\columnwidth]{figures/RMSE_vrs_SF_1038.pdf}
        \caption{}
        \end{subfigure}
      \begin{subfigure}{8cm}
       \includegraphics[width=1\columnwidth]{figures/RMSE_vrs_SF_remcomGrid64.pdf}
          \caption{}
        \end{subfigure}
      \fi
    \caption{Comparison with state-of-the-art alternatives
      \secondrev{when the training and testing maps were obtained from
        the Wireless InSite data set and $Q_t=10$. (a) Square area of
        side $100$ m with $N_y=N_x=32$; (b) square area of side $200$
        m with $N_y=N_x=64$.} \secondrev{The autoencoder that produces the
        blue curve in (b) has two additional Conv2D and
        Conv2DTranspose layers, code length $\latentnum=256$, and uses
        $T=1.25\cdot 10^6$.}}
    \label{f:comparisonwinsite}
  \end{figure*} 
  
      \end{myitemize}
 \subsubsection{Wireless Insite Data Set}
  \label{sec:experimentspowermapwinsite} 
 \begin{myitemize}
   \myitem\cmt{training}To investigate how the proposed network would perform in a 
   real-world setup, training uses the Wireless Insite data set 
   in combination with the technique in
   Sec.~\ref{sec:realworldreal}, where the sets $\Omega_{t,q}^{(I)}$ and $\Omega_{t,q}^{(O)}$
are drawn from $\Omega_t$ uniformly at random
   without replacement with    $\vert\Omega_{t,q}^{(I)}\vert=\vert\Omega_{t,q}^{(O)}\vert=1/2\vert
   \Omega_t\vert$, $q=1,\ldots,Q_t$,  and $Q_t=10~\forall t$.
    \myitem\cmt{Figures}Figs.~\ref{f:comparisonwinsite}\secondrev{a} and \secondrev{\ref{f:comparisonwinsite}b} show the RMSE
    as a function of $\vert \Omega \vert$ for the proposed scheme and
    competing alternatives \secondrev{with two area sizes}. By the performance degradation of all \firstrev{five}
    approaches relative to Fig.~\ref{f:comparison}, it follows that
    estimating real maps is more challenging than estimating maps in
    the Gudmundson data set. The performance gap is increased, where
    the proposed approach now performs roughly \firstrev{100} \secondrev{ and 90 \% in Figs.~\ref{f:comparisonwinsite}a and~\ref{f:comparisonwinsite}b}, respectively, better than the next
    competing alternative. \firstrev{As in Fig.~\ref{f:comparison}, the algorithm
    in~\cite{ding2016cellular} is not displayed  due to a high RMSE.} 
 \end{myitemize}
 \subsubsection{\firstrev{Hybrid Training}}
  \label{sec:experimentspowertransfer}
   
  \begin{figure}[t]
\centering
\if\singlecol1
\includegraphics[width=0.50\columnwidth]{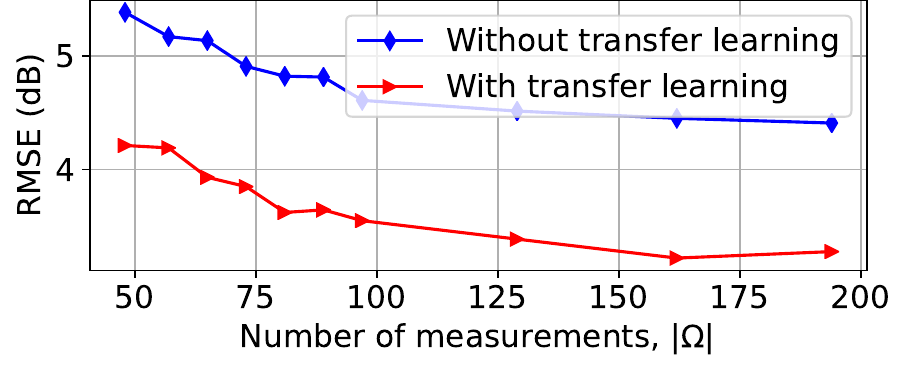}
 \else
 \includegraphics[width=1\columnwidth]{figures/rmse_trLear.pdf}
 \fi
\caption{\firstrev{Performance gain due to hybrid training using $T=10^3$ maps
  with $Q_t=30$ from the Wireless InSite data set. The initial weights used
  for the blue curve were drawn at random whereas those for the red curve
  were obtained by training over the Gudmundson data set.}}
\label{f:rmse_trLear}
\end{figure}

  \firstrev{The performance of the hybrid training scheme from
    Sec.~\ref{sec:hybridtraining} is analyzed in
    Fig.~\ref{f:rmse_trLear}.  As expected,  transfer learning yields a significant performance
    improvement.}

  \subsection{Deep Neural Network Design}
 \label{sec:netdesign} 
 \begin{myitemize}%
  \myitem\cmt{overview}This section justifies the main design
  decisions regarding the proposed network.  \myitem\cmt{number of
    parameters appr. same}To unveil the influence of each
  architectural aspect,  the number
  of convolution and convolution-transpose filters is adjusted so that
  the total number of parameters of the neural network $N_{\pars}$
  remains approximately the same.
  
  \myitem\cmt{Experiment 3 \ra RMSE vs codelength}The first step is to
  justify the choice of an autoencoder structure. To this end, Fig.~\ref{f:rmse_vs_code} complements the toy example in Fig.~\ref{f:motivatingex} by plotting  the RMSE as a function of the code length $\latentnum$ under two setups 
  \begin{myitemize}%
  \myitem\cmt{pathloss only}with pathloss propagation and fixed transmit power:
  \myitem\cmt{2 cases}i) Noisy \emph{inputs}, noiseless \emph{targets} in the training phase, and noisy \emph{inputs}, noiseless \emph{targets} in the testing phase. This corresponds to training as a denoising autoencoder; see Sec.~\ref{sec:realworldsynthetic}. ii) Noisy \emph{inputs} and \emph{targets} in the training, and noisy \emph{inputs}, noiseless \emph{targets} in the testing. This models how a neural network trained over real data estimates a true map. 
  \myitem\cmt{Figure}Note that the irregular behavior of the curves
  for $\latentnum > 5$ owes to the fact that each $\latentnum$
  corresponds to a different network, and therefore a different
  training process, including the initialization.  As observed, the
  RMSE remains roughly constant for $\latentnum > 5$, which
  demonstrates that the spectrum maps in this scenario lie close to a
  low-dimensional manifold. This justifies the autoencoder
  structure. Besides, training as a denoising autoencoder offers a
  slight performance advantage, yet it is only possible with synthetic
  data; see Sec.~\ref{sec:realworldsynthetic}. When other propagation
  phenomena such as shadowing need to be accounted for, 
  $\latentnum>5$ is however required.
    
    \if\singlecol1
    \begin{figure}[t!]
    \centering
     \begin{minipage}{0.48\textwidth}
      \includegraphics[scale=0.55]{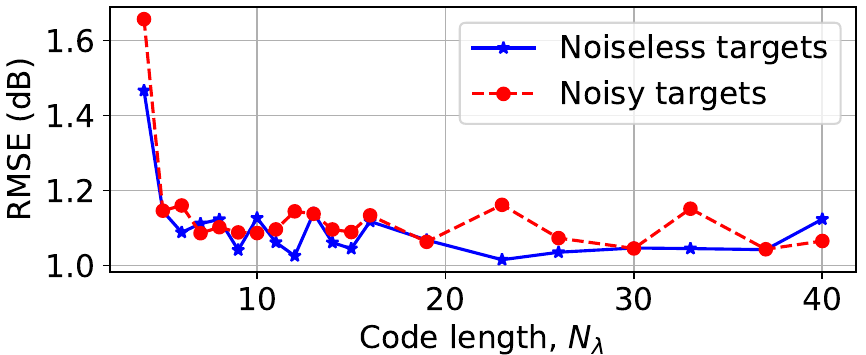}
       \caption{RMSE as a function of the code length $\latentnum$, $|\Omega|=104$. 
    }
      \label{f:rmse_vs_code}
      \end{minipage}\hfill
   \begin{minipage}{0.48\textwidth}
         \includegraphics[scale=0.55]{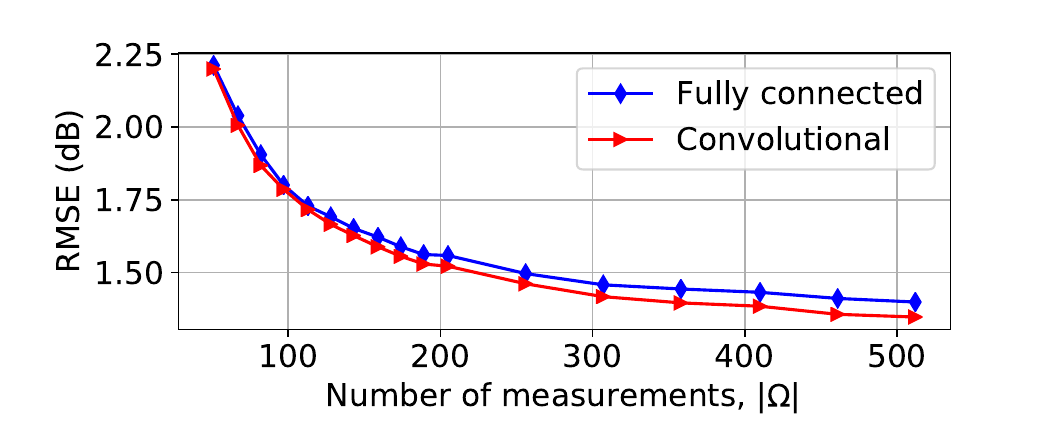}
          \caption{RMSE of the proposed network as a function of $|\Omega|$ for two types of output layers for the encoder. 
    }
    \label{f:rmse_vs_sf_layertypes}
   \end{minipage}
    \end{figure}
      \else
      \begin{figure}[t!]
    \centering
      \includegraphics[width=1\columnwidth]{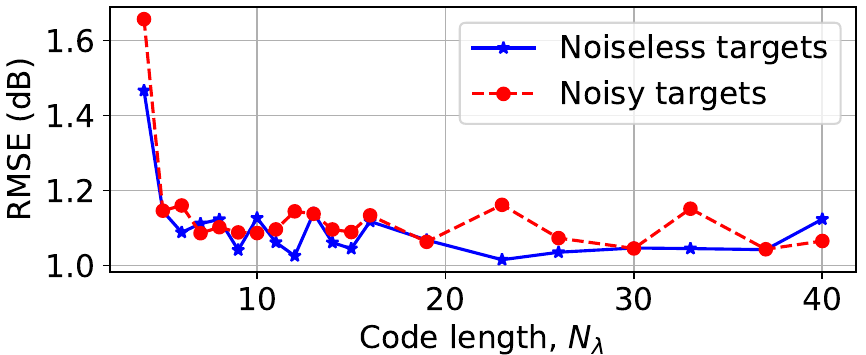}
      \caption{RMSE as a function of the code length $\latentnum$, $|\Omega|=104$. 
    }
     \label{f:rmse_vs_code}
    \end{figure}
     \begin{figure}[t!]
    \centering
      \includegraphics[width=1\columnwidth]{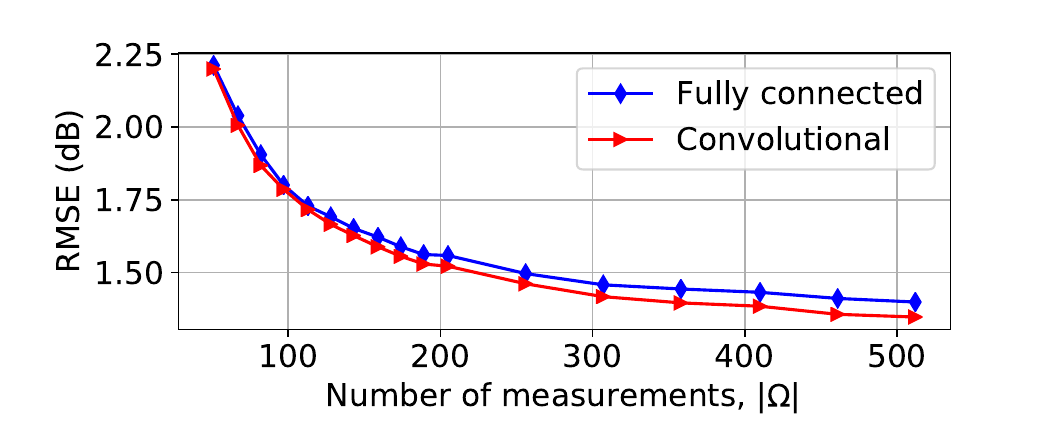}
    \caption{RMSE of the proposed network as a function of $|\Omega|$ for two types of output layers for the encoder. 
    }
    \label{f:rmse_vs_sf_layertypes}
  \end{figure} 
      \fi
      \end{myitemize}

  \myitem\cmt{Experiment 2 \ra fully conv. vs dense layer net.}A
  second design consideration is whether the last layer of the encoder
  should be convolutional or fully connected. In the former case, the
  code would capture shift-invariant features, whereas greater
  flexibility is allowed in the latter case. This dilemma is
  ubiquitous in deep learning since convolutional layers constitute a
  special case of fully connected layers. The decision involves the
  trade-off between flexibility and information that can be learned
  with a finite number of training examples. This is investigated in 
  \begin{myitemize}%
  \myitem\cmt{Figure}Fig.~\ref{f:rmse_vs_sf_layertypes}, which shows
  the RMSE as a function of the number of measurements $|\Omega|$ for
  these two types of layers. As observed, in the present case, fully
  convolutional autoencoders perform slightly better. Besides, they
  accommodate inputs of arbitrary $N_x$ and $N_y$. For
  these reasons, the proposed architecture is fully convolutional.
  \end{myitemize}

  \myitem\cmt{Experiment 1 \ra RMSE vs num. of layers}Two more design
  decisions involve the number of layers $\layernum$ and the choice of
  the activation functions.
  \begin{myitemize}%
  \myitem\cmt{Figure}Fig.~\ref{f:rmse_vs_numlayers} shows the RMSE as
  a function of $\layernum$ with LeakyReLU and PReLU
  activations~\cite{he2015delving}, where the latter generalize the
  former to allow training the leaky parameter. Recall that the number of
  neurons per layer is adjusted to yield approximately the same number
  of training parameters for all $\layernum$. Thus, this figure
  embodies the trade-off between the number and complexity of the
  features extracted by the network as well as the impact of
  overfitting.  As observed, the best performance in this case is achieved around
  $\layernum =26$ layers. Both activations yield roughly the same
  RMSE, yet the PReLU outperforms the LeakyReLU for shallow
  architectures. 
    \begin{figure}[t!]
    \centering
    \if\singlecol1
        \includegraphics[scale=0.6]{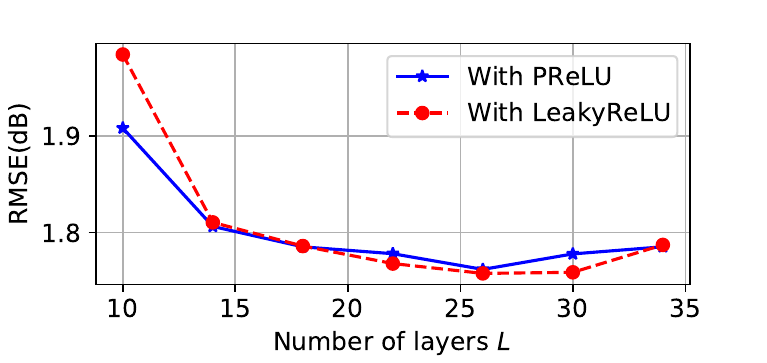}
      \else
      \includegraphics[width=1\columnwidth]{figures/RMSE_vrs_num_layers_1033.pdf}
      \fi
    \caption{Estimation RMSE of the proposed approach as a function of the number of layers $\layernum$ of the autoencoder for two different activation functions, $|\Omega|=300$.
    }
    \label{f:rmse_vs_numlayers}
  \end{figure}
   \end{myitemize}

    \begin{figure*}[t!]
    \centering
    \begin{subfigure}{8cm}
\centering\includegraphics[width=7cm,height=11.5cm]{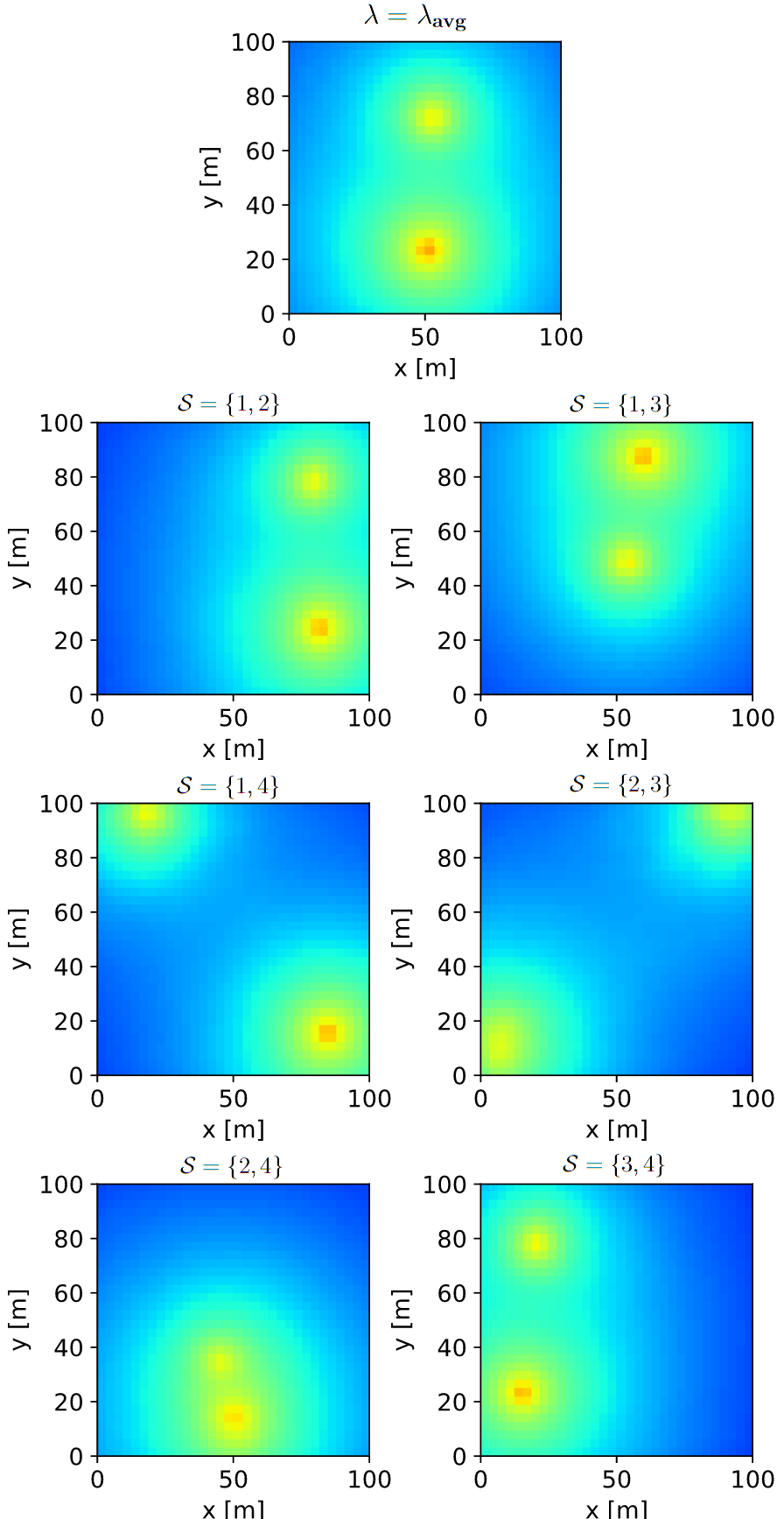}
\caption{}
\end{subfigure}%
    \begin{subfigure}{8cm}
\centering\includegraphics[width=7cm,height=11.5cm]{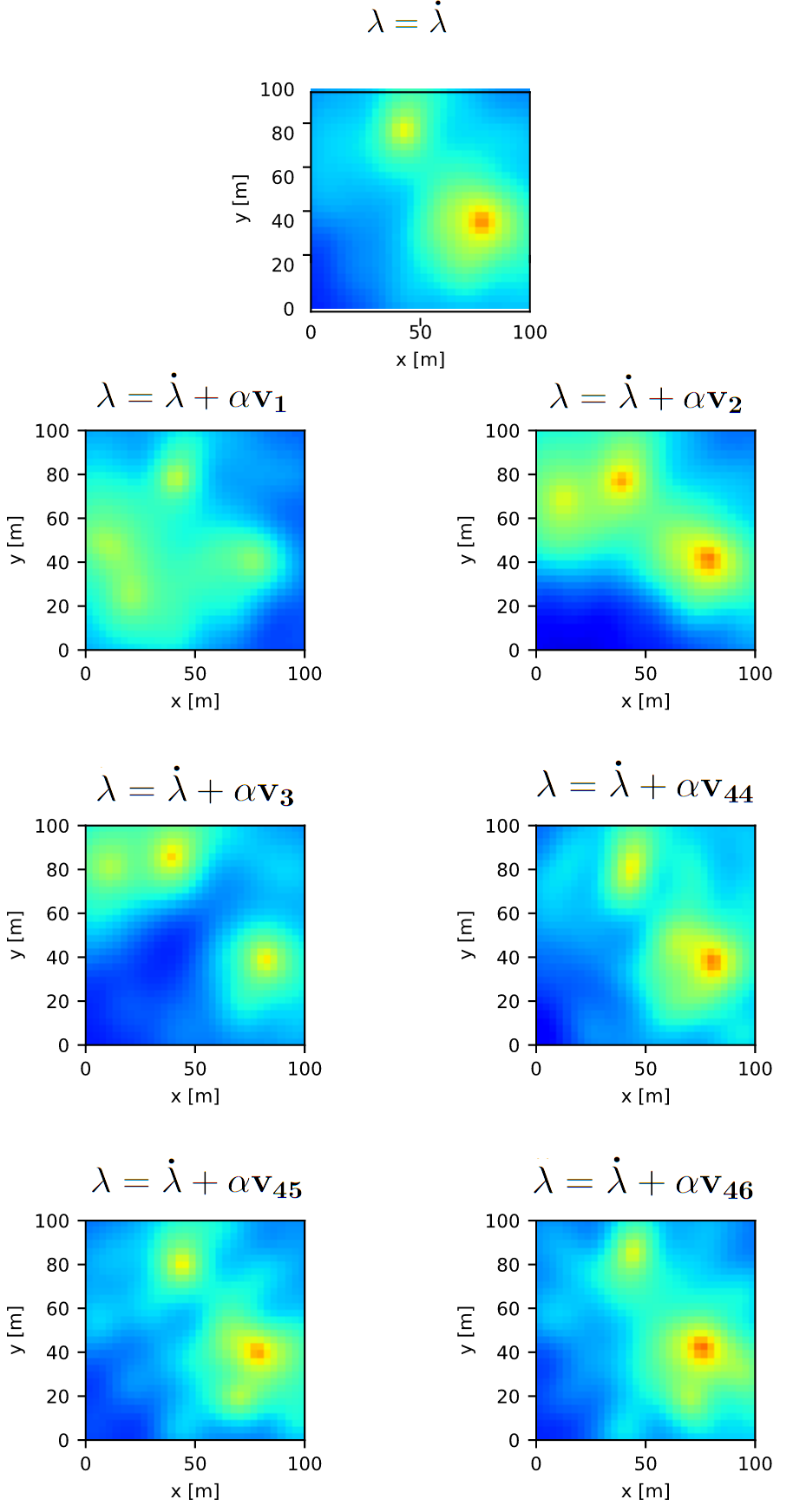}
\caption{}
\end{subfigure}%
    \caption{Decoder outputs of autoencoder architectures with different code length and trained with different data sets: (a) $\latentnum=4$ with maps from the free-space propagation model, (b) $\latentnum=64$ with maps from the Gudmundson data set, $\alpha=10$.}
    \label{f:visualiz_maps}
  \end{figure*}
    \begin{figure}[t!]
\begin{center}
\if\singlecol1
\includegraphics[width=0.51\columnwidth]{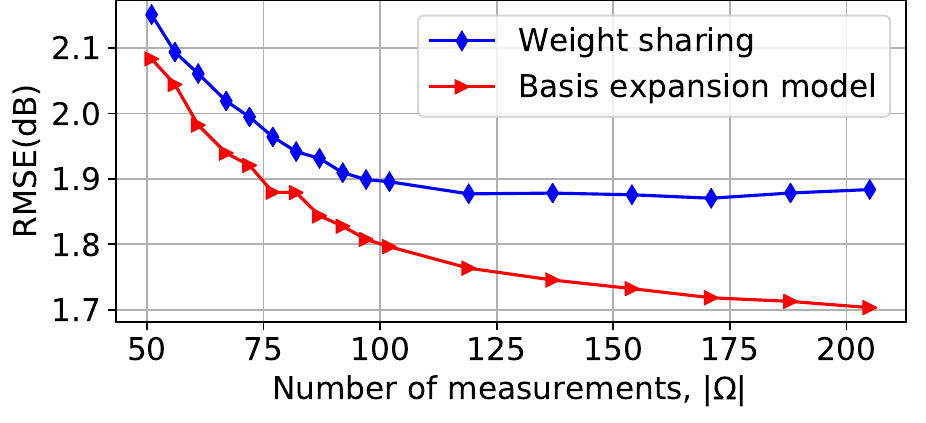}
\else
\includegraphics[width=1\columnwidth]{figures/rmse_w_sharing_bem.pdf}
\fi
\caption{\secondrev{Map estimation RMSE with and without prior information  in the frequency domain.}} 
\label{f:rmse_w_sharing_bem}
\end{center} 
\end{figure}
     \begin{figure*}[t!]  
    \centering
     \if\singlecol1
         \includegraphics[width=1\columnwidth, height=8cm]{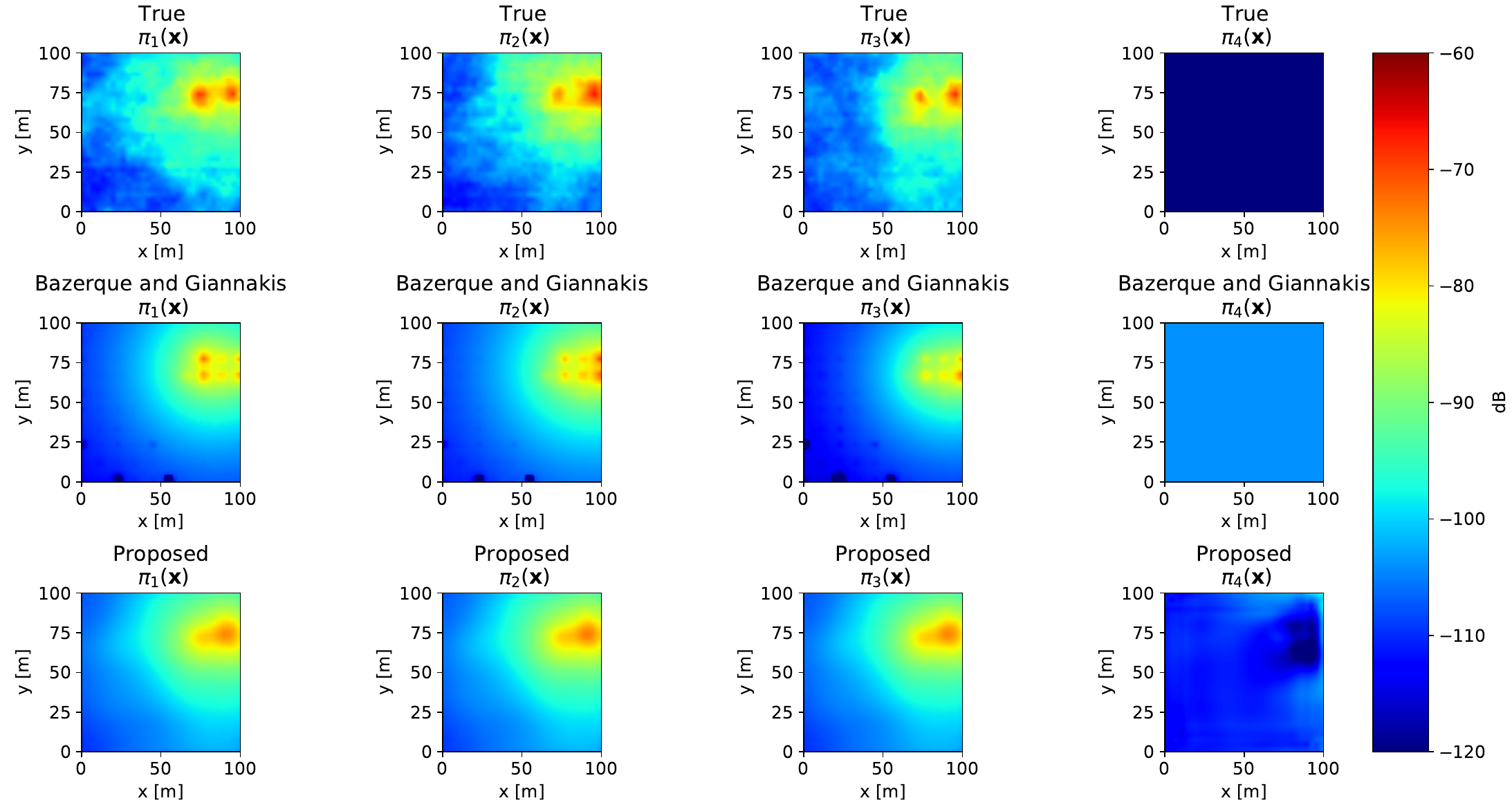}
      \else
      \includegraphics[scale=0.5]{figures/True_and_Estimated_bcoeffs_1035_gaussian_base_and_noise_base_205measurs.pdf}
      \fi
 
    \caption{Maps of the true and estimated cofficients $\{ \truecoeffs_\baseind(\bm x)\}_{\baseind=1}^{\basenum}$ over $\mathcal{X}$, $\basenum=4$. 
    }
    \label{f:bemcoefficients}
  \end{figure*} 

\subsection{Feature Visualization}
\cmt{Overview}Although neural networks are mainly treated as black
boxes, some visualization techniques offer interpretability of the
features that they extract and, therefore,  shed light on the nature of the
information that is learned. 
\myitem\cmt{Experiment 4 \ra visualization figures}To this end, the
next experiment depicts the decoder output when different latent vectors
$\code\in \rfield^{\latentnum}$ are fed at its input.

\begin{myitemize}%
  \myitem\cmt{path loss case \ra vary 2 variables at a time}First, an instance of the proposed
  autoencoder with $\latentnum=4 $ is trained with a dataset of
  $T_{\latentnum} = 3\cdot 10^3$ maps generated using the free-space
  propagation model with two sources transmitting with a fixed
  power. Since these maps only differ in the x and y coordinates of the
  sources, they form a $4$-dimensional manifold.  Applying the
  encoder to those maps yields $\{\code_t\}_{t=1}^{T_{\latentnum}}$.
  \begin{myitemize}%
    \myitem\cmt{Figure \ra mean code}The top panel of Fig.~\ref{f:visualiz_maps}a depicts the output of the trained decoder when ${\code}= \code_{\text{avg}}$, where $\code_{\text{avg}}\define (1/T_{\latentnum}) \sum_{t} \code_t$. As expected, the decoder reconstructs a map with two sources.
    
  \myitem\cmt{vary latent vars.}The code $\code$ acts as the coordinates of a map in the learned manifold. To study this manifold, the output of the decoder is depicted for different values of these coordinates.
  Specifically, each
  of the remaining panels in Fig.~\ref{f:visualiz_maps}a corresponds
  to a value of $\code = \check\code$ with
  $[\check{\code}]_k=[\code_{\text{avg}}]_k - [\code_{\text{std}}]_k$
  if $k\in \mathcal{S}$ and $[\check{\code}]_k=[\code_{\text{avg}}]_k$
  otherwise, where $[\code_{\text{std}}]_k\define\sqrt{
    \sum_t\left([\code_t]_k-[\code_{\text{avg}}]_k \right)^2 /
    T_{\latentnum}}$ and the set $\mathcal{S}$ is indicated in the panel
  titles. It can be observed that moving along the manifold
  coordinates produces maps of the kind in the training
  set. 
  
     \end{myitemize}   
  \myitem\cmt{shadowing case \ra using code covariance matrix}These
  panels focus on path loss. To understand how shadowing is
  learned,  an instance of the proposed autoencoder with code $\code\in
  \rfield^{\latentnum},\,\latentnum=64$, is trained with the Gudmundson data set.
  \begin{myitemize}%
\myitem\cmt{random code}The top panel of Fig.~\ref{f:visualiz_maps}b
depicts the output of the trained decoder for ${\code}= \dot \code$,
where $\dot \code$ was chosen uniformly at random among
$\{\code_t\}_{t=1}^{T_{\latentnum}}$. As expected, the decoder
reconstructs a map with two sources and the effects of shadowing are
noticeable.
\myitem\cmt{perturbation}To introduce perturbations in
this code along directions that are informative to different extents,
let $\latentcovmat \define (1/T_{\latentnum})( \latentvarmat -
\code_{\text{avg}}\bm 1^\top) ( \latentvarmat - \code_{\text{avg}}\bm
1^\top)^\top \in \rfield ^{\latentnum \times \latentnum}$ denote the
sample covariance matrix of the $T_{\latentnum}$ training codes, where
$\latentvarmat\define[\code_1, \ldots, \code_{T_{\latentnum}}]$.
The latent vectors are set to
${\code}=\dot \code + \alpha \bm v_i$, where $\alpha$ is a fixed
constant and $\bm v_i$ is the $i$-th principal eigenvector of
$\latentcovmat$.  The remaining panels of
Fig.~\ref{f:visualiz_maps}b show the map estimates  for $i = 1, 2, 3, 44, 45, 46 $ and
$\alpha=10$. As anticipated, changes along the directions of high
variability yield maps with markedly different shadowing patterns.
The opposite is observed by moving along directions of lower
variability,  where the reconstructed maps are roughly similar to
the one in the top panel.
 \end{myitemize} 
  \end{myitemize}  
%
   

  \if\singlecol1
    \begin{figure}[t!]
    \centering
     \begin{minipage}[t]{0.48\textwidth}
      \includegraphics[scale=0.50]{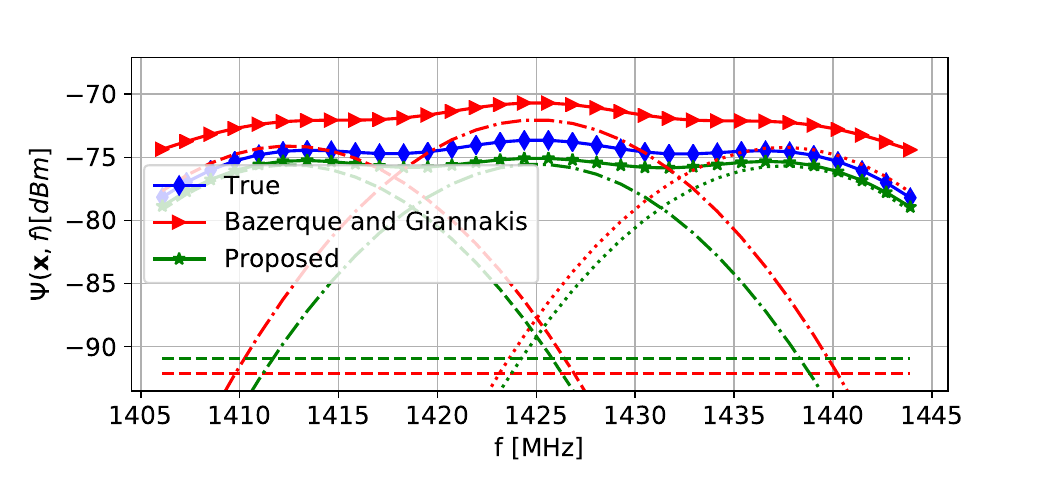}
       \caption{PSD reconstruction at a random location $\bm x \in\mathcal{X}$ where the basis expansion model uses Gaussian functions. The non-continuous red (green) curves represent the products $\estimatecoeffs_\baseind(\bm x)\bases_{\baseind}(f)$ estimated by the competing (proposed) algorithm.}
      \label{f:psdreconstructiongaussian}
      \end{minipage}\hfill
   \begin{minipage}[t]{0.45\textwidth}
          \includegraphics[scale=0.5]{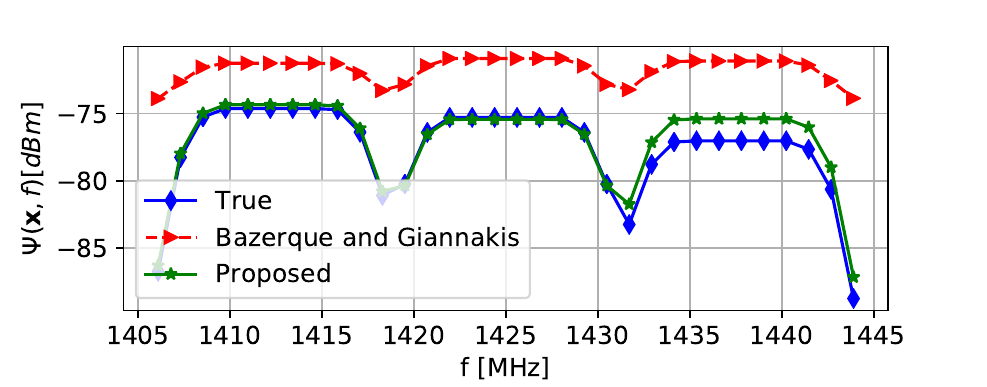}
          \caption{PSD reconstruction at $\bm x \in\mathcal{X}$ with a signal basis formed by using raised-cosine functions. 
    }
    \label{f:psdreconstructionraisedc}
   \end{minipage}
    \end{figure}
      \else
      \begin{figure}[t!]
    \centering
      \includegraphics[width=1\columnwidth]{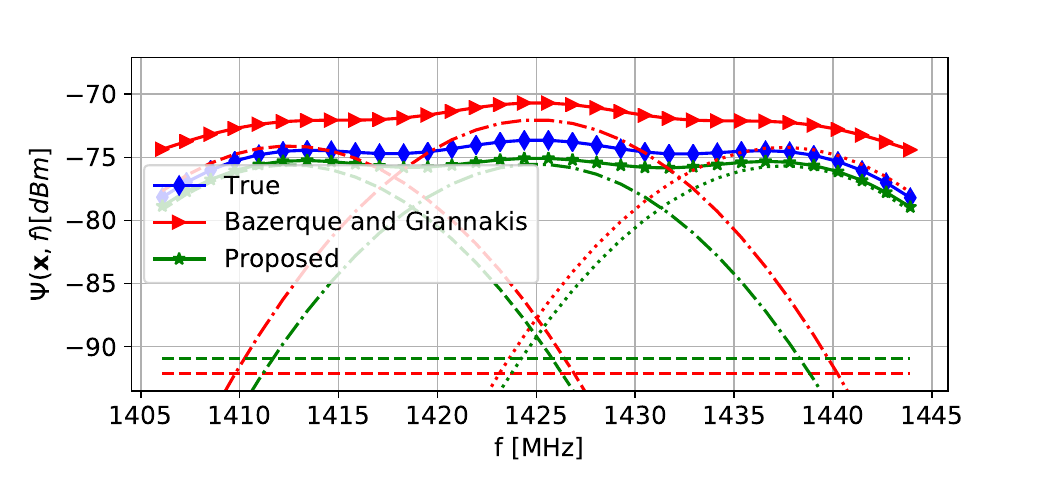}
      \caption{PSD reconstruction at a random location $\bm x \in\mathcal{X}$ where the basis expansion model uses Gaussian functions. The non-continuous red (green) curves represent the products $\estimatecoeffs_\baseind(\bm x)\bases_{\baseind}(f)$ estimated by the competing (proposed) algorithm. 
    }
     \label{f:psdreconstructiongaussian}
    \end{figure}
     \begin{figure}[t!]
    \centering
      \includegraphics[width=1\columnwidth]{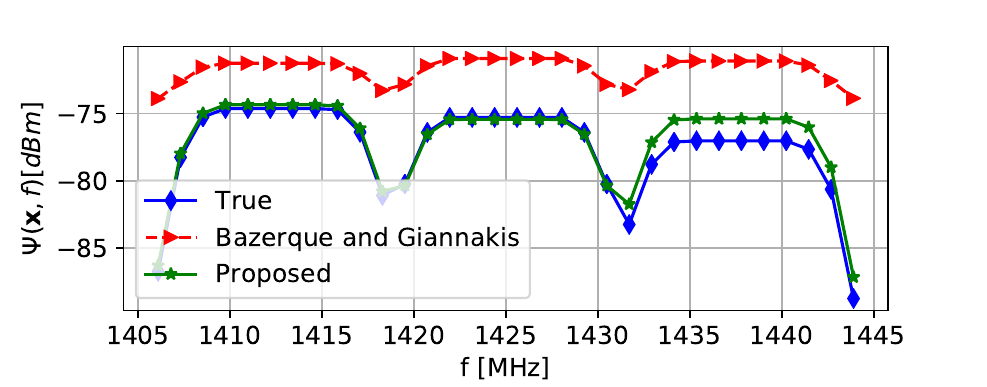}
    \caption{PSD reconstruction at $\bm x \in\mathcal{X}$ with a signal basis formed by using raised-cosine functions.}
    \label{f:psdreconstructionraisedc}
  \end{figure} 
      \fi

    \subsection{PSD Cartography}
    \label{sec:experimentspsdmap}
\cmt{overview}This section provides empirical support for  the
approach proposed in Sec.~\ref{sec:completionbem} for PSD cartography.
To this end, each sensor samples the received PSD  at $N_f=32$
uniformly spaced frequency values in the band of interest. 
\cmt{competing algor.}The performance of the proposed method is
compared with that of the non-negative  Lasso algorithm
in~\cite{bazerque2010sparsity} with regularization parameter
$10^{-11}$, which yields approximately the best RMSE. To
improve its performance, this algorithm was extended to assume that
the noise power is the same at all sensors.
 \begin{figure}  
    \centering
    \if\singlecol1
         \includegraphics[scale=0.56]{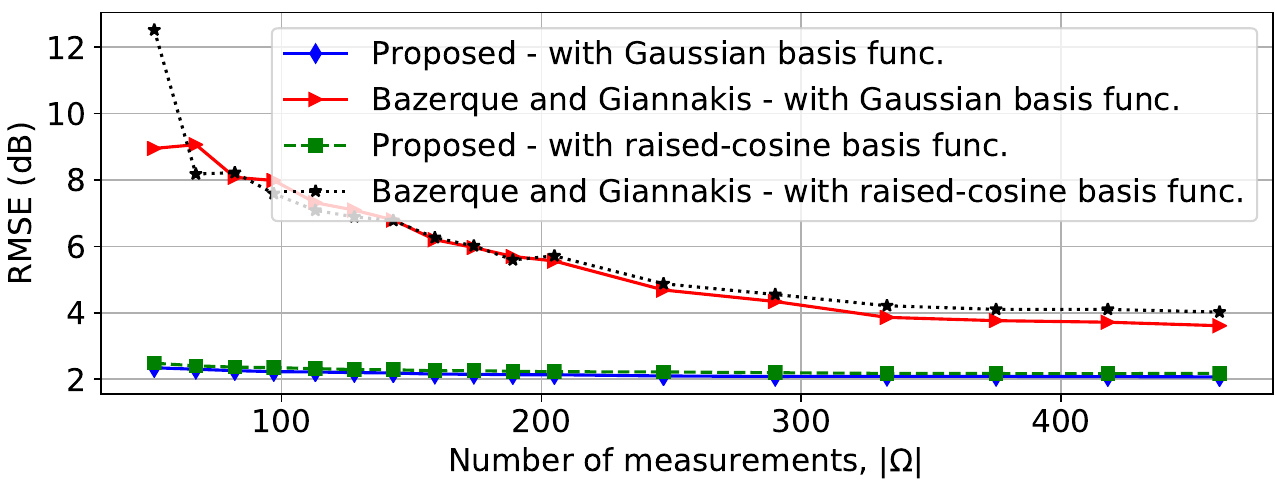}
      \else
      \includegraphics[width=1\columnwidth]{figures/RMSE_vrs_SF_1035.pdf}
      \fi
    \caption{Performance comparison of the proposed scheme with that of the algorithm in~\cite{bazerque2010sparsity}. 
    }
    \label{f:comparisonpsdgud}
  \end{figure} 
     \begin{figure}[t!]
    \centering
     \if\singlecol1
        \includegraphics[scale=0.55]{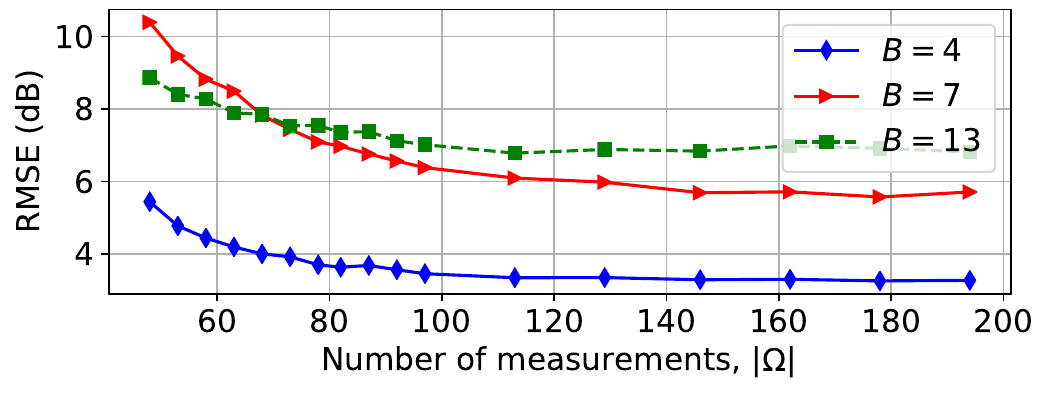}
      \else
      \includegraphics[width=1\columnwidth]{figures/RMSE_vrs_SF_1039.pdf}
      \fi
    \caption{Map estimate RMSE of the proposed scheme for PSD cartography. The training and testing maps were obtained  from the Wireless InSite data set, $Q_t=5$.}
    \label{f:rmsepsdwinsite}
  \end{figure}
        \subsubsection{Gudmundson Data Set}%
          \label{sec:experimentspsdmapgudmund}               
        \begin{myitemize}%
    \myitem\cmt{overview}The first part of this section
    \secondrev{evaluates the approaches in
      Sec.~\ref{sec:exploitstruc}} and assesses \secondrev{the performance of the proposed algorithms}
     using the training strategies in
    Sec.~\ref{sec:realworldsynthetic} when the training and testing
    maps were obtained from the Gudmundson data set.
    \myitem\cmt{basis functions}The $B-1=3$ \emph{signal} basis
    functions are uniformly spaced across the band, whereas a fourth
    constant basis function is introduced to model \emph{noise}; see
    Sec.~\ref{sec:completionbem}.  Two types of signal basis functions
    are investigated: Gaussian radial basis functions with standard
    deviation 5 MHz and raised-cosine functions with roll-off factor
    0.4 and bandwidth 10 MHz.  \myitem\cmt{noise psd}The noise basis
    function is scaled to yield $\upsilon(\bm x, f)=\upsilon$, where
    $\upsilon$ is a uniform random variable between $-100$ and $-90$
    dBm/MHz.
         
        \myitem\cmt{Figure \ra 2 poss comparison}\secondrev{
          Fig.~\ref{f:rmse_w_sharing_bem} demonstrates the superiority
          of the completion autoencoder when the output layers
          proposed in Sec.~\ref{sec:completionbem} are utilized
          relative to the weight-sharing scheme from
          Sec.~\ref{sec:noprior}. The reason is that the latter does
          not exploit prior information in the frequency domain.  This
          output layer will be used in all remaining experiments.  }
    
    \myitem\cmt{Figures}The top row of Fig.~\ref{f:bemcoefficients}
    portrays the maps of the true coefficients $\{
    \truecoeffs_\baseind(\bm x)\}_{\baseind=1}^4$ over $\mathcal{X}$;
    the second and last rows show their estimates with both schemes
    when $\vert \Omega \vert=512$. Visually, the proposed scheme
    produces better estimates despite the fact that it does not
    exploit the fact that the noise power is the same at all
    sensors. To demonstrate the reconstruction quality of the proposed
    scheme, Figs.~\ref{f:psdreconstructiongaussian}
    and~\ref{f:psdreconstructionraisedc} show the true and estimated
    PSDs at a random location  $\bm x \in \mathcal{X}$. As observed,
    the PSD estimate produced by the proposed scheme follows the true
    PSD more closely compared to the one produced by the competing
    algorithm. A quantitative comparison is provided in
    Fig.~\ref{f:comparisonpsdgud}, which shows the RMSE as a function
    of the number of measurements $\vert \Omega \vert$. As observed,
    the proposed method outperforms the competing approach with
    significant margin for small $\vert \Omega \vert$.
    
    \end{myitemize}       
    \subsubsection{Wireless Insite Data Set}
     \label{sec:experimentspsdwinsite}
    \begin{myitemize}
    \myitem\cmt{overview}The second part of this section evaluates the performance of the proposed scheme using the training approach in Sec.~\ref{sec:realworldreal}, where the sets $\Omega_{t,q}^{(I)}$ and $\Omega_{t,q}^{(O)}$
are drawn from $\Omega_t$ uniformly at random
   without replacement with    $\vert\Omega_{t,q}^{(I)}\vert=\vert\Omega_{t,q}^{(O)}\vert=1/2\vert
   \Omega_t\vert$, $q=1,\ldots,Q_t$,  and $Q_t=5~\forall t$. The
   training and testing maps were obtained  from the Wireless InSite
   data set.
   \myitem\cmt{basis functions}The transmit PSD is generated with the
    raised-cosine functions described in
   Sec.~\ref{sec:experimentspsdmapgudmund}. 
        \myitem\cmt{noise psd}The noise PSD  is set to $\upsilon(\bm x, f)=\upsilon$, where $\upsilon$ is a uniform random variable between $-180$ and $-170$ dBm/MHz.  
    \myitem\cmt{Figure}Fig.~\ref{f:rmsepsdwinsite} shows the RMSE of
    the proposed method as a function of the number of measurements
    $\vert \Omega \vert$. Because of the high RMSE of the competing
    approach~\cite{bazerque2010sparsity}  (possibly in part due to the reasons in Remark~\ref{rem:dbfitting}), its performance is not shown
    on the figure. 
As observed, the proposed scheme yields a  low RMSE
    in this realistic scenario which emulates training with real
    measurements. 
    \end{myitemize}
      
    \myitem\cmt{run-time comparison}\secondrev{To conclude, it is
      worth stressing the computational efficiency of the trained
      autoencoder when estimating radio maps, since a simple forward
      pass is required. This contrasts with state-of-the-art alternatives,
      which generally require the application of iterative algorithms
      or the inversion of large matrices. As a reference, using our
      implementations for power map estimation in \secondrev{Fig.~\ref{f:comparisonwinsite}a on an Intel Core i7-6820HQ CPU}, the proposed scheme takes $25 \cdot 10^{-3}$
      seconds, whereas the kriging, multi-kernel, GPR, and KNN methods
      respectively need $10^{-1}$, $1.25$, $3 \cdot 10^{-1}$, and $4
      \cdot 10^{-1}$ seconds.  For PSD map estimation, the
       run-time of the proposed algorithm is around $2.12 \cdot
      10^{-1}$ seconds whereas that of the  algorithm
      in~\cite{bazerque2010sparsity} is $6.43$ seconds.}

\end{myitemize}

\section{\firstrev{Conclusions and Discussion}}
\label{sec:conclusion}
\begin{myitemize}%
\myitem\cmt{Summary}Data-driven radio map estimation has been proposed to learn the spatial structure of propagation phenomena such as shadowing, reflection, and diffraction. \firstrev{Provided that sufficiently realistic data sets are available,} learning such structure from past measurements yields estimators that require fewer measurements \firstrev{than state-of-the-art alternatives to attain a target
performance. Motivated by the observation that radio maps lie close to
a low-dimensional manifold embedded in a high-dimensional space, a
deep completion network with an encoder-decoder architecture was
proposed to estimate PSD maps. Extensive numerical experiments with
two datasets reveal that the resulting scheme can attain an RMSE in the
order of 2 dB, significantly better than existing approaches.}
\myitem\cmt{limitations}\firstrev{The price to be paid for such an improved
performance is the need for a training data set and increased
computational demands compared to traditional ``interpolation''
approaches. Besides, deep neural networks typically require large data
sets and obtaining the latter may be costly when one relies on
measurements or ray tracing algorithms. However, this is a common
limitation of any reasonable data-driven approach and can be
alleviated by means of transfer learning.}
\myitem\cmt{Future work}Future work will include mapping other channel metrics such as channel-gain with alternative network architectures. 
\end{myitemize}%

\if\editmode1 
\onecolumn
\printbibliography
\else
\bibliography{\bibfilenames}
\fi
\end{document}